\documentclass[twocolumns,traditabstract]{aa}
\usepackage{graphicx}
\usepackage{epsfig}
\usepackage{lscape}
\usepackage{graphicx}
\usepackage{amsmath}
\usepackage{natbib}
\usepackage[english]{babel}
\usepackage[titletoc,title]{appendix}
\usepackage{txfonts}
\usepackage{times}

\begin{document}
\title{The Relative Growth of Black Holes and of the Stellar Components of Galaxies }
\author{N. Menci, F. Fiore, A. Bongiorno, A. Lamastra \inst{1}
  }
\institute{INAF - Osservatorio Astronomico di Roma, via di Frascati 33, 00040 Monte Porzio Catone, Italy
}

\abstract{
Recent observations indicate that the  mass of Supermassive Black Holes (SMBHs) correlate differently with different galaxy stellar components. 
Comparing such observations with the results of ``ab initio" galaxy formation models can provide  insight on the mechanisms leading to the 
growth of SMBHs. Here we use a state-of-the-art semi-analytic model of galaxy formation to investigate the correlation of the different galaxy stellar components with the mass of the central SMBH. The stellar mass in the disc, in the bulge, and in the pseudo-bulge of galaxies is related to quiescent star formation, to galaxy interactions, and to the loss of angular momentum following disc instabilities, respectively. Consistently with recent findings, we find that while the predicted bulge masses are tightly correlated with the SMBH masses, the correlation between the latter and the galactic discs shows a much larger scatter, in particular when bulgeless galaxies are considered. In addition, we obtain that the predicted masses of pseudo-bulges shows little or no-correlation with the masses of SMBHs. We track the histories of merging, star formation, and SMBH accretion to investigate the physical processes at the origin of such findings within the context of cosmological models of galaxy formation. Finally, we discuss the effects of variations of our assumed fiducial model on the results. 
}
\authorrunning{N. Menci et al.}
\maketitle
\vspace{-0.2cm}
\section{Introduction}

 The discovery of relations between the mass $M_{BH}$ of super-massive Black Holes (SMBHs) and the properties of the spheroidal component of the host galaxies,  like the  stellar mass  $M_{sph}$ (Kormendy \& Richstone 1995, Magorrian et al. 1998; Marconi \& Hunt 2003; Häring \& Rix 2004) and the  stellar velocity dispersion (e.g., Ferrarese \& Merritt 2000; Gebhardt et al. 2000; Merritt \& Ferrarese 2001; Tremaine et al. 2002; Greene \& Ho 2006; Graham 2007; Gültekin et al. 2009), has shown that the growth of BHs must be tightly connected to the processes governing  the evolution of galaxies. Stimulated by the above observations, a 
 number of theoretical studies (see e.g. Kauffmann \& Haehnelt 2000; Monaco, Salucci \& Danese 2000; Granato et al. 2004; Hopkins et al. 2006; Fontanot et al. 2006; Fanidakis et al. 2012; Hirschmann et al. 2012, 2014; Lamastra et al 2010; Menci et al. 2003, 2014) have included the growth of SMBHs in the framework of cosmological galaxy formation models. Most of them  assumed mergers and/or disc instabilities as triggers for BH accretion, and related the effectiveness of each accretion event to the galaxy properties (like the fraction of cold available gas, the merging rate, or the mass of the central accreting BH) computed 
through ``ab initio" semi-analytic models (SAMs) of galaxy formation (see Somerville \& Dav\'e 2015 for a review). These approaches showed that $M_{BH}-M_{sph}$ relations matching the observations can result in the framework of hierarchical galaxy formation models. Inclusion of sub-grid prescriptions for SMBH accretion into  hydrodynamical simulations also allowed
to obtain $M_{BH}-M_{sph}$ relations matching the observations, either due to self-regulation following AGN feedback (in the case 
of Bondi accretion strongly dependent on $M_{BH}$, see e.g., Di Matteo et al. 2005; Dubois et al. 2012) or to the evolution of galaxy gas content and merging rates controlling BH feeding (see, e.g., Angl\'es-Alc\'azar, Ozel, Dav\'e 2013 in the case of torque-limited growth of SMBHs). 


In recent years, refined observational studies of the relation between the SMBH mass and the stellar content of galaxies  (e.g.,  Bennert et al. 2011; Graham et al. 2011; Kormendy \& Bender 2011; Kormendy et al. 2011; McConnell et al. 2011; Sani et al. 2011; Beifiori et al. 2012; Graham \& Scott 2013; McConnell \& Ma 2013) allowed to investigate  the role of different physical processes like dry mergers (see, e.g, Graham \& Scott 2013),  bars (see Hu 2008; Graham \& Li 2009 ),  or disc instabilities (see, e.g., Cisternas et al. 2011), 
 in shaping the co-evolution of BHs and host galaxies, with the aim of pinning down the basic mechanisms that trigger the growth of BHs in relation with the  building up of the different stellar populations of the host galaxies. For example, pseudo-bulges (spheroids with disc-like exponential profiles or rotational kinematics) have been reported to lie below the main relation defined by regular bulges and ellipticals (Hu 2008; Greene et al. 2010; Shankar et al. 2012),  or not to correlate at all with the BH masses (Kormendy, Bender, Cornell 2011). Such a  behaviour would have deep implications, since in the common view 
 classical bulges are formed through violent processes such as mergers (Toomre \& Toomre 1972; Toomre 1977; Schweizer 1990; Kormendy 2012 and references therein; Avile-Reese, Zavala,  Lacerna 2014) while pseudo-bulges are believed to be formed through internal evolution in isolated galaxies, involving disc instabilities,  angular momentum transfer and ensuing inflows (see Wyse, Gilmore \& Franx 1997; Kormendy \& Kennicutt 2004; Athanassoula 2005; see also Gadotti 2009), though the formation of classical bulges through disc instabilities has been considered in some models 
 (see, e.g., Perry, Eke, Frenk 2009; De Lucia et al. 2011) as we discuss in the final section. In addition, recent studies showed that massive BHs can be hosted in bulgeless galaxies as shown by the presence of AGN in bulgeless galaxies (see Ho 2008; Kormendy \& Ho 2013 for reviews). 

Overall, the above results suggest a scenario in which
 the different stellar populations (those residing in discs, in pseudo-bulges and in bulges) show a different degree of correlation with the BH mass
 (see Kormendy \& Ho 2013 for a review; for recent results based on megamaser measurements see L\"asker et al. 2016). It is then interesting to investigate how such an emerging picture fits into galaxy formation models. 
 This is particularly relevant in view  of the different mechanisms that - in present models - drive the formation of the different stellar component of galaxies (see Somerville \& Dav\'e 2015; Silk \& Mamon 2012 for recent reviews): quiescent, secular conversion of cold disc gas into stars for the disc component; major merging for classical bulges; and disc instabilities for the pseudo-bulge component. Thus, comparing the observed BH – host galaxy correlations for the different stellar components with results of galaxy formation models may help in understanding which of the processes involved in galaxy evolution are directly related to the growth of SMBHs. 
 
Here we use a state-of-the-art SAM of galaxy formation to investigate the correlation of the different galaxy stellar components with the mass of the central SMBH. The model includes the physics of the galactic gas, the quiescent mode of star formation converting the cold disc gas into stars on timescales $\sim $ Gyr, the impulsive mode of star formation corresponding to starbursts triggered by interactions, and the star formation related to inflows following disc instabilities (see Hopkins 2011)  implemented as described in previous papers (Menci et al. 2014; Gatti et al. 2015). 
The AGN feedback is computed following the expansion of the shock front through the interstellar medium as described in detail in Menci et al. (2008). In our fiducial model, the BH growth is triggered mainly by galaxy interactions (including fly-by), but we shall also discuss a scenario where BH growth is entirely driven by inflows due to disc instabilities. In our model, although the AGN feedback plays a role in controlling the gas cooling and the star formation in massive galaxies, the BH-stellar mass relation is mainly controlled by how gas feeding is related to the galaxy properties and evolution. These, in turn,  relate differently with the different stellar components. The comparison with the observed relations will allow to study the connection of the different processes driving the formation of the different stellar components (secular gas conversion, interactions, disc instabilities) with the growth of SMBHs. 
\vspace{-0.2cm}
\section{The Semi-Analytic Model}

We base on the semi-analytic model described in Menci et al. (2014) and Gatti et al. (2015), to which we refer for details; here we recall its key points. 
The merging trees of dark matter haloes are generated through a Monte Carlo procedure adopting the merging rates given by the Extended Press \& Schechter formalism (see Bardeen, Bond, Kaiser, Szalay 1991; Lacey \& Cole 1993; Bower 1991) starting from redshift $z=15$. We assume 
a Cold Dark Matter (CDM) power spectrum of perturbations, and we adopted a total matter density parameter $\Omega_0=0.3$, a baryon density parameter $\Omega_b=0.04$, a Dark Energy density parameter $\Omega_{\Lambda}=0.7$, and a Hubble constant $h=0.7$ in units of 100 km/s/Mpc. For each dark matter halo included into a larger halo we computed the dynamical friction process, to 
determine whether it will survive as a satellite, or sink to the centre to increase the mass of the central dominant galaxy; binary interactions (fly-by and merging) among satellite sub-halos are also described by the model. The gas in each  halo is  initially set to have a density given by the universal baryon fraction and to be at the virial temperature. We then compute its cooling due to atomic processes. The cooled gas fraction settles into a rotationally supported disc with mass $M_c$, disc radius $R_d$ and disc circular velocity $V_d$ computed as in Mo, Mao \& White (1998). The gas is converted into stars through three different channels described in sect. 2.2: quiescent star formation, gradually converting the gas into stars over long timescales $\sim 1$ Gyr; starbursts following galaxy interactions, occurring on timescales $\lesssim 100$ Myr, associated to BH feeding; internal disc instabilities triggering loss of angular momentum resulting into gas inflows toward the centre, therefore feeding star formation and BH accretion. 

The energy released by the Supernovae associated to the total star formation returns a fraction of the disc gas into the hot phase,  providing the feedback needed to prevent overcooling. An additional source of feedback is provided by the energy radiated by the Active Galactic Nuclei (AGN) which corresponds to the active accretion phase of the central BH described below in sect. 2.1; the detailed description of our implementation of the AGN feedback is given in Menci et al. (2008). Finally, the luminosity - in different bands - produced by the stellar populations of the galaxies are computed by convolving the star formation histories of the galaxy progenitors with a synthetic spectral energy distribution, which we take from Bruzual \& Charlot (2003) assuming a Salpeter IMF. The model includes the tidal stripping of the part of the stellar content of each satellite galaxy  for which we adopt exactly the treatment introduced by  Henriques \& Thomas (2010). We remind that this is based on the computation of the tidal radius for each satellite galaxy; this is identified as the distance from the satellite centre at which the radial forces acting on it cancel out (King 1962; Binney \& Tremaine 1987; see also Taylor \& Babul 2001). These forces are the gravitational binding force of the satellite, the tidal force from the central halo and the centrifugal force. In the simple approximation of nearly circular orbits and of an isothermal halo density profile,  such a radius can be expressed as $r_t\approx \sigma_{sat}\,r_{sat}/\sqrt{2}\,\sigma_{halo}$, where $\sigma_{sat}$ and $\sigma_{halo}$ are the velocity dispersions of the satellite and of the halo, respectively, and $r_{sat}$ is the halocentric 
radius of the satellite, that we computed in our Monte Carlo code during its orbital decay to the centre due to dynamical friction. For each satellite galaxy, the material outside this radius is assumed to be disrupted and becomes a diffuse stellar component in the host halo. Stellar stripping could play a role in producing outliers in the $M_{BH}-M_*$ relation, as shown in the EAGLE hydrodynamical simulations 
by Barber et al. (2016).  

\subsection{Black Hole Accretion}

The semi-analytic model includes BH growth from primordial seeds. These are assumed to originate from PopIII stars with a mass $M_{seed}=100\,M_{\odot}$  (Madau \& Rees 2001), and to be initially present in all galaxy progenitors. Our results are insensitive to the specific value of the seed mass as long as $M_{seed}\lesssim 10^5\,M_{\odot}$.
We consider two BH feeding modes: accretion triggered by galaxy interactions and internal disc instabilities. 
These are described in detail in Menci et al, (2014) and Gatti et al. (2015), and briefly recalled below.

a) BH accretion triggered by interactions. The interaction rate $\tau_r^{-1}=n_T\,\Sigma (r_t,v_c,V)\,V_{rel} (V)$ for galaxies with relative velocity $V_{rel}$ and number density $n_T$ in a common DM halo determines the probability for encounters, 
either fly-by or  merging, through the corresponding cross sections $\Sigma$ given in Menci et al. (2002; 2014). The fraction of
gas destabilized in each interaction corresponds to the loss $\Delta j$ of orbital angular momentum $j$, and depends on the mass ratio of the merging partners $M'/M$ and on the impact factor $b$ as follows: 
\begin{equation}\label{fdest}
f\approx \frac{1}{2}\,
\Big|{\Delta j\over j}\Big|=
\frac{1}{2}\Big\langle {M'\over M}\,{R_d\over b}\,{V_d\over V_{rel}}\Big\rangle\, ,
\end{equation}
We assume that in each interaction $1/4$ of the destabilized gas mass $f\,M_c$ feeds the central BH, while the remaining fraction feeds the circumnuclear starbursts (Sanders \& Mirabel 1996). Thus, the BH accretion rate is given by
\begin{equation}
\label{macc_ID}
 {dM_{BH}\over dt}={1\over 4}{f\,M_c\over \tau_d}  
\end{equation}
where the time scale $\tau_{d}=R_d/V_d$  is assumed to be the crossing time of the  galactic disc. .

b) BH accretion induced by disc instabilities. As in other SAMs (see, e.g., Lo Faro et al. 2009; De Lucia et al. 2011; Hirschmann et al. 2012), we assume these to arise  in  galaxies with disc mass exceeding $M_{crit} =  {v_{max}^2 R_{d}/ G \epsilon}$ 
(Efstathiou et al. 1982) with $\epsilon=0.75$, where $v_{max}$ is the maximum circular velocity associated to each halo (Mo, Mao, \& White 1998). 
Such a criterion strongly suppresses the probability for disc instabilities to occur not only in massive, gas-poor galaxies, but also in 
dwarf galaxies characterized by small values of the gas-to-DM mass ratios.
The instabilities induce loss of angular momentum resulting into strong inflows that we compute following the 
description in Hopkins et al. (2011), recast and extended as in Menci et al. (2014). The resulting BH accretion rate  (in units of 
$M_{\bigodot} yr^{-1}$) is

\begin{equation*}
\label{hopkins}
\frac{dM_{BH}}{dt}  \approx 
{\alpha \,
f_d^{4/3}\over 1+2.5\,f_d^{-4/3}(1 + f_0/f_{gas}) } \, \times 
\left( \frac{M_{BH}}{10^8 M_{\bigodot}}\right)^{1/6}
\left( \frac{M_d}{10^9 M_{\bigodot}}\right)
\end{equation*}
\begin{equation}
{\rm with}~~~~
f_0 \approx 0.2 f_d^2 \left[ \frac{M_d}{10^9 M_{\bigodot}}\right]^{-1/3}
~~~  f_{gas} \equiv {M_{gas} \over M_d}
\end{equation} 
Here $M_{BH}$ is the central black hole (BH) mass,  $f_d$ is the total disc mass fraction, and $M_d$ and $M_{gas}$ the disc and the gas mass calculated at the reference radius $R_0 = 100$ pc  (writing the above equation in terms of different reference radius $R_0$ would not change the 
resulting accretion rate, see Menci et al. 2014).
The constant $\alpha$ depends on the exact form of the assumed disc surface density profile and to the shape of the Schmidt star formation law in the inner region (see Hopkins et al. 2011). Present observational uncertainties in such quantities yield an allowed range $1\leq \alpha\leq 5$. In the following we shall take $\alpha = 5$;  this value approximates that corresponding to our assumed  Kennicutt-Schmidt law 
relation $\dot M_*\propto M_c/\tau_d$ with a precision better than 5\% (see eq. A14 in Menci et al. 2014).  Note that the expression in eq. 3 
assumes the persistence of a stable disc (indeed, it is derived in the WKB approximation, see Hopkins \& Quataert 2011, Menci et al. 2014) 
and hence does not hold in our major merging events. In addition, the 
accretion rate in eq. 3 has a mild dependence on the BH mass, but depends very strongly on the disc mass fraction and on the gas mass fractions ($f_d$ and $f_{gas}$). Thus,  we expect disc instabilities to be effective in gas rich systems which underwent a limited number of merging events during their past history. 

In our model, at any given time BHs are allowed to grow via both channels. 
As discussed in Menci et al. (2014) the build-up of the final BH masses is dominated by the interaction-driven accretion (see sect. 2.3 below and fig. 2). However, inflows produced by disc instabilities contribute to the star formation of the galaxy and drive the growth of the pseudo-bulge, as we discuss below. 
 
\subsection{Star formation}

Our model includes three modes of star formation: 

i)  The quiescent mode, corresponding to the gradual conversion of disc gas into  stars with a rate $\dot M_*\propto M_c/\tau_d$ directly related to the  
  disc time defined above. The proportionality constant is given by the measured local Schmidt-Kennicutt law yielding a conversion time $\tau_*(z=0)=1$ Gyr (e.g., Santini et al. 2014). The stars formed through this channel are assigned to the galactic disc.

ii) The impulsive mode, corresponding to starbursts triggered by galaxy interactions: the interaction rate is that given at point a) in sect. 2.1. The fraction of gas converted into stars corresponds to $3/4\,f$ where the destabilized gas fraction $f$ is that given in eq. (2.1). Following Hopkins et al. (2009) we assume that in mergers with mass ratio $\mu\geq 0.2$ a fraction $1-f_{gas}$ of the disc mass is transferred to the bulge, while the stars formed in minor mergers and fly by are assigned to the disc. 
The resulting distribution of bulge-to-total ($B/T$) stellar mass ratio has been tested against observations in Menci et al. (2014). Note, however, that our following results  will not change appreciably if we take the canonical assumption that all disc stars are transferred into the bulge during major mergers. 

iii) Disc instabilities.These arise when the disc mass exceeding $M_{crit} =  {v_{max}^2 R_{d}/ G \epsilon}$, as discussed in Sect. 2.1b. The corresponding loss of angular momentum induces nuclear star formation. Since the model in Hopkins et al. (2011) assumes an equilibrium between the mass inflow and star formation, the latter reads $\dot M_*=A_*\,\dot M_{BH,DI}$ (see Menci et al. 2014 for the computation), where $\dot M_{BH,DI}$ is the BH accretion rate in eq. 3. The exact value of the proportionality constant $A_*\geq 10^2$ depends on the detailed radial profile of the disc potential and of the gas surface density. In Menci et al. (2014) we obtain a value  $A_*\approx 100$ in the nuclear  
region of the galaxy, while the contribution from the outer disc regions may yield an additional $\sim$ 50\% the nuclear value in the case of an exponential density profile. In the following we shall keep the value $A_*\approx 100$ as  our fiducial choice; larger values would result into a slight shift of the pseudo-bulge masses, and would not change our main results (see below and Sect. 3). 
  The stars formed through the disc instability mode are assigned to a pseudo-bulge component, according to a widely adopted view (see, e.g., Combes et al. 1990; Raha et al. 1991; Debattista et al. 2006; Kormendy \& Kennicutt 2004; Athanassoula et al. 2005; see also Méndez-Abreu  2014; Obreja et al. 2013) also implemented in other SAMs (see, e.g., Fontanot et al. 2015a). However, we note that alternative explanations (that we shall not consider here) have been proposed; e.g., Eliche-Moral et al. (2011) suggested that pseudo-bulges might also be created by the secular accretion of low-density satellites into the main galaxy, while   Keselman \& Nusser (2012) propose a merger origin for pseudo-bulges. In our implementation,  such a 
  star formation mode requires large fraction of gas left available in the galaxy, so that the gas mass can exceed the mass threshold $M_{crit}$ (sect. 2.1) for the onset of disc instabilities, and to allow for large values of the disc and gas fractions $f_d$ and $f_{gas}$ entering the accretion rate in eq. 3.

 Each model galaxy is allowed to have both a bulge and a pseudo-bulge. When comparing with 
data concerning  pseudo-bulge we shall consider model galaxies where such stellar component dominates over the classical bulge component.

\subsection{The Relative Growth of the BH and Total Stellar  Content of Galaxies}

The model allows to follow the BH accretion (sect. 2.1) and  star formation (sect. 2.2) in all the progenitors along the 
merger history of any given final galaxy. In fig. 1 we show how the growth histories of $M_*$ and $M_{BH}$ are distributed among the progenitors of two selected final (z=0) galaxies in our SAM (with mass $M_*\approx 10^{11}\,M_{\odot}$). The merger trees result from our Monte Carlo realizations described above (see the beginning of Sect. 2), and include all the dynamical processes affecting the evolution of sub-halos.  Each branch represents a progenitor of the final galaxy at the redshift indicated by the vertical axis, and the size and color correspond to the fractional mass (normalized to the final mass) of the stellar  (left panel) and BH  (right panel) component of each progenitor, as results from the full computation of our SAM. 

\begin{center}
\vspace{-0.4cm}
\hspace{-0.4cm}
\scalebox{0.2}[0.2]{\rotatebox{-0}{\includegraphics{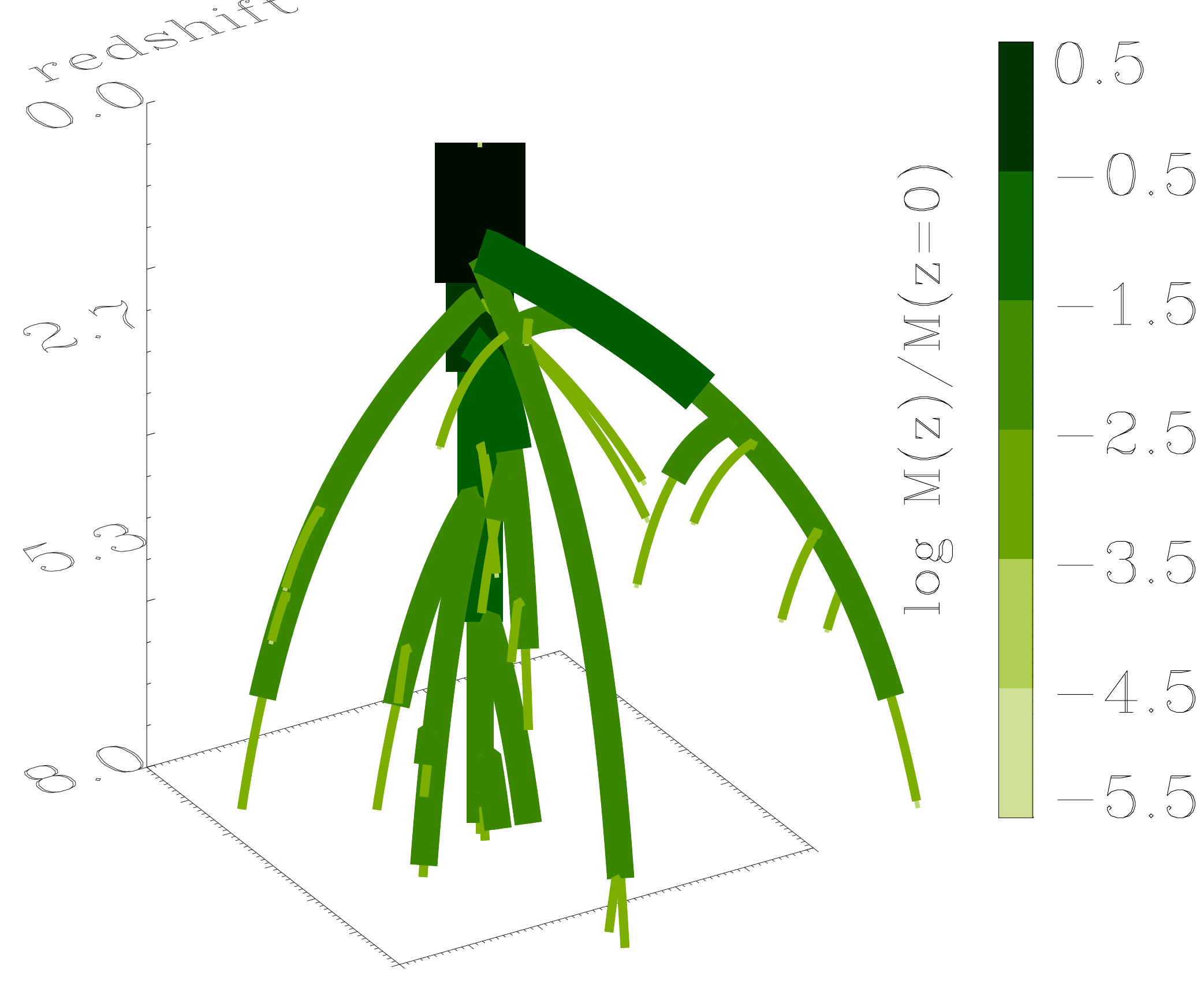}}}
\hspace{0.cm}
\vspace{-0.2cm}
\scalebox{0.2}[0.2]{\rotatebox{-0}{\includegraphics{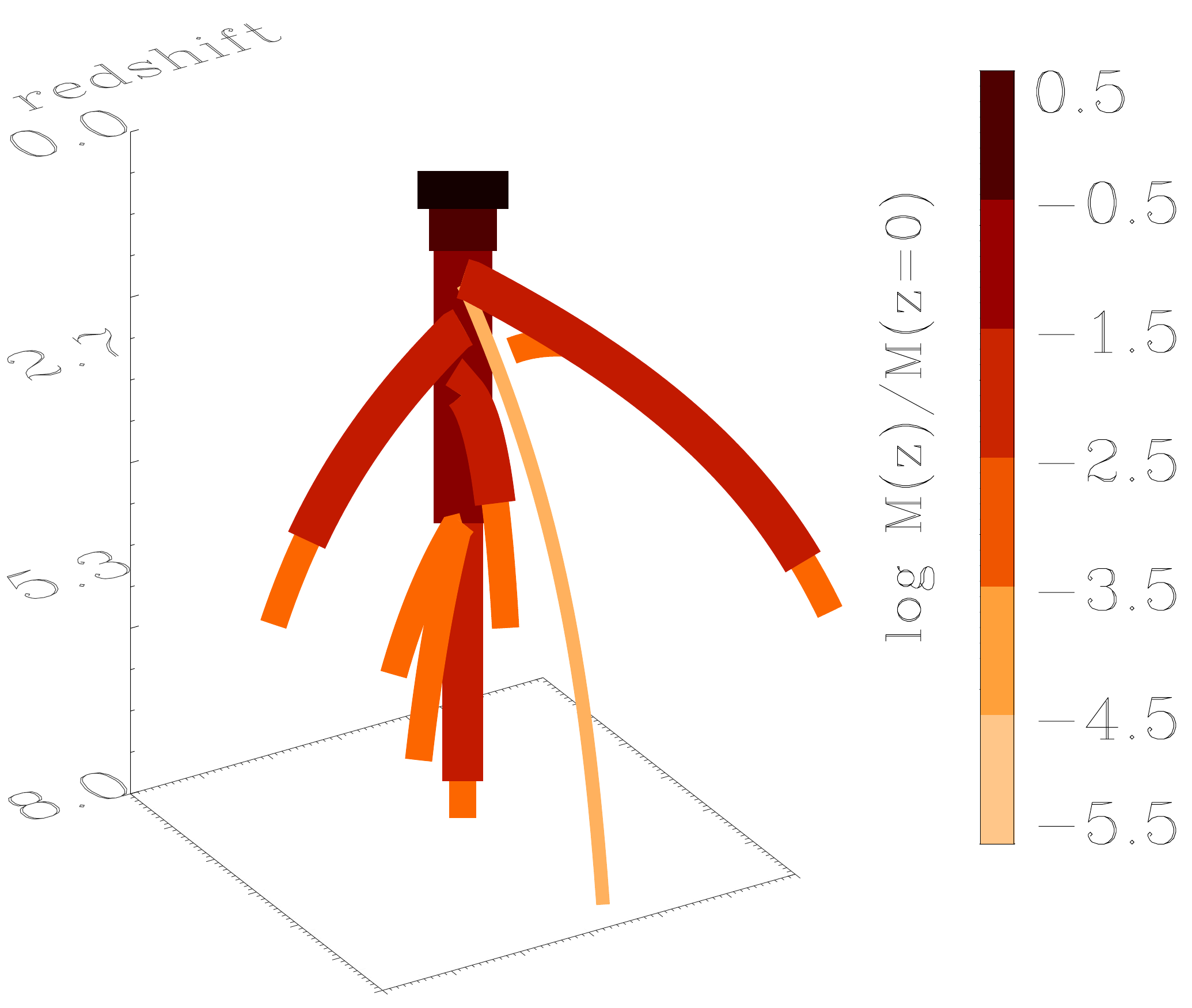}}}
\vspace{-0.8cm }
\end{center}
\begin{center}
\hspace{-0.4cm}
\scalebox{0.2}[0.2]{\rotatebox{-0}{\includegraphics{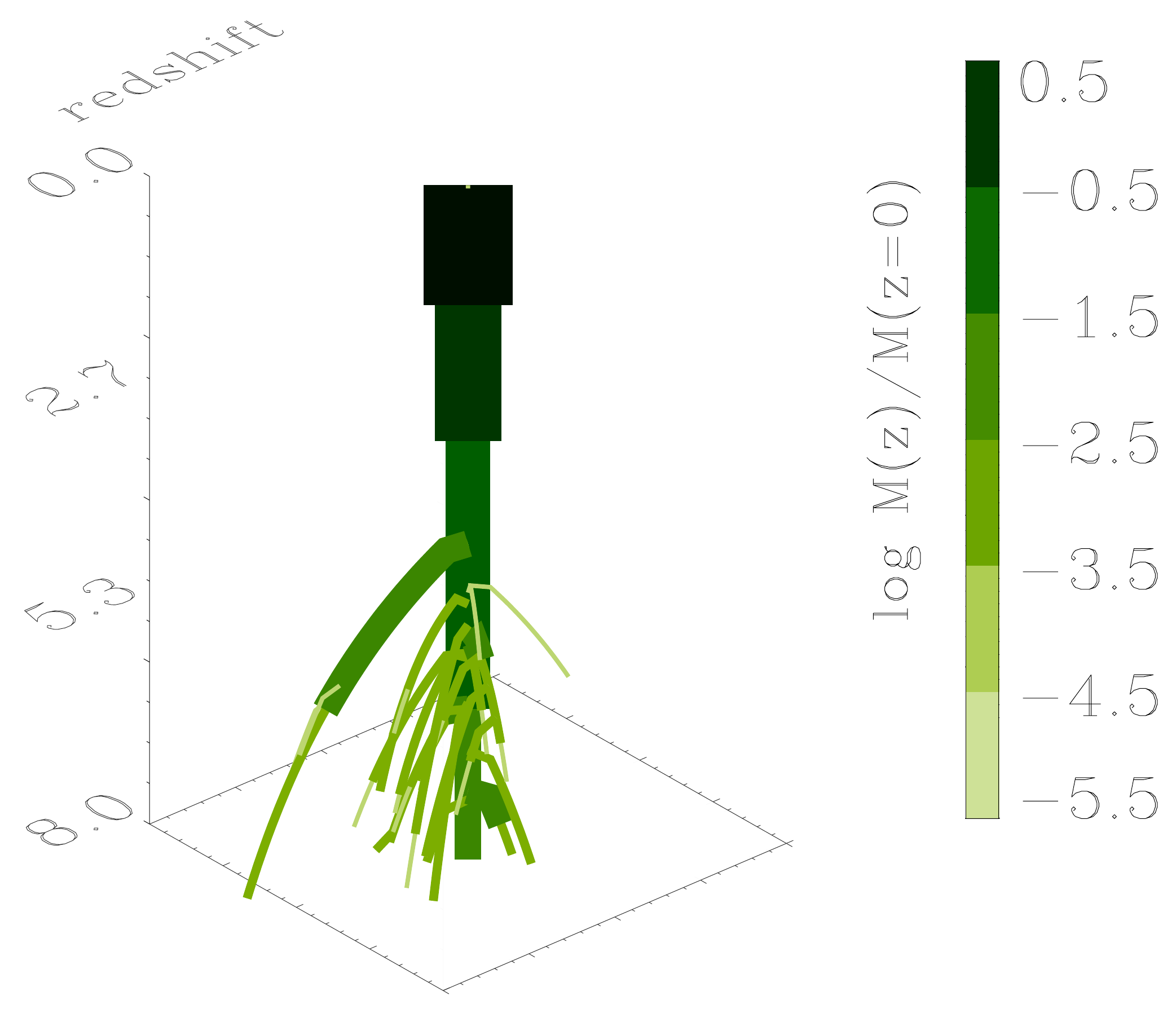}}}
\hspace{0.cm}
\vspace{-0.2cm}
\scalebox{0.2}[0.2]{\rotatebox{-0}{\includegraphics{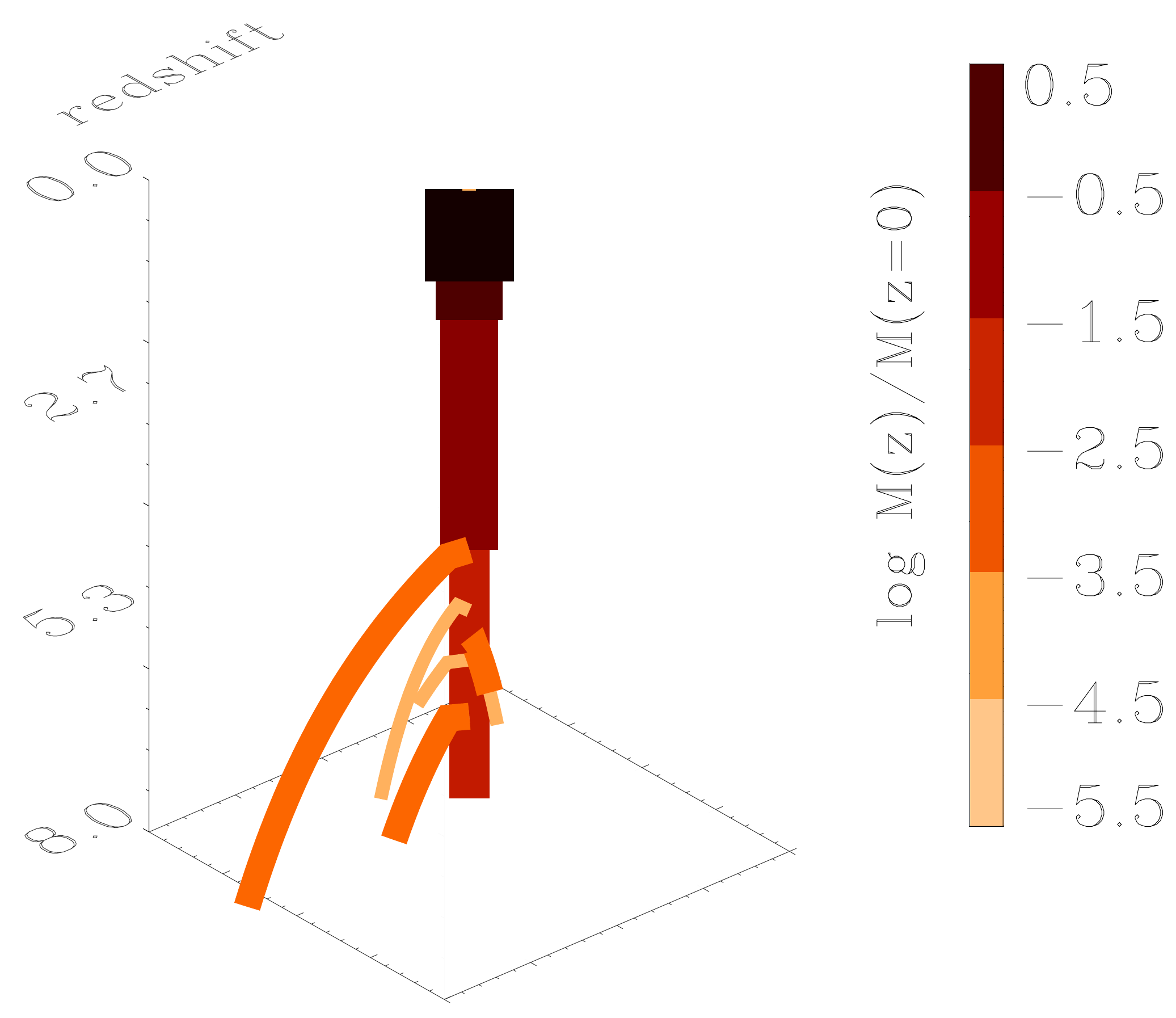}}}
\end{center}
\vspace{-0.4cm }
 {\footnotesize 
Fig. 1. Selected Merging Trees. 
Each branch represents a progenitor of the final galaxy at the redshift indicated by the vertical axis. The spatial distribution of branches in the 
$x$-$y$ plane is arbitrary, chosen to separate branches in order to obtain a better graphical visualization.
Color and size of each branch correspond to the stellar mass (left panels) and BH mass (right panels) associated to the progenitor galaxies.
 The corresponding logarithmic mass scale in units of the final mass is shown in the colorbars. To avoid overcrowding in the figure, we showed only branches with masses larger than $10^{-5}$ the final (z=0) value. Thus, accretion of the small lumps containing the seed BH masses in not 
represented. The first history leads to final a galaxy with B/T=0.5, the second to a final elliptical galaxy with B/T=0.9.
\vspace{0.1cm}}

We note that, while at high redshift the stellar mass is distributed among numerous progenitors (since it is contributed also by the quiescent mode not depending on galaxy merging or interactions), the contributions to the final BH mass are mainly  in the main progenitor and in the few main branches where most of the merging activity is concentrated.  The latter takes place mainly at high redshifts $\gtrsim 3$. 

The statistical effect of all the growth histories results into the final BH mass distribution shown in fig. 2a and in the 
local relation between the BH mass and the total stellar mass in fig. 2b. 

\vspace{-0.2cm}
\hspace{-0.8cm}
\scalebox{0.26}[0.26]{\rotatebox{-0}{\includegraphics{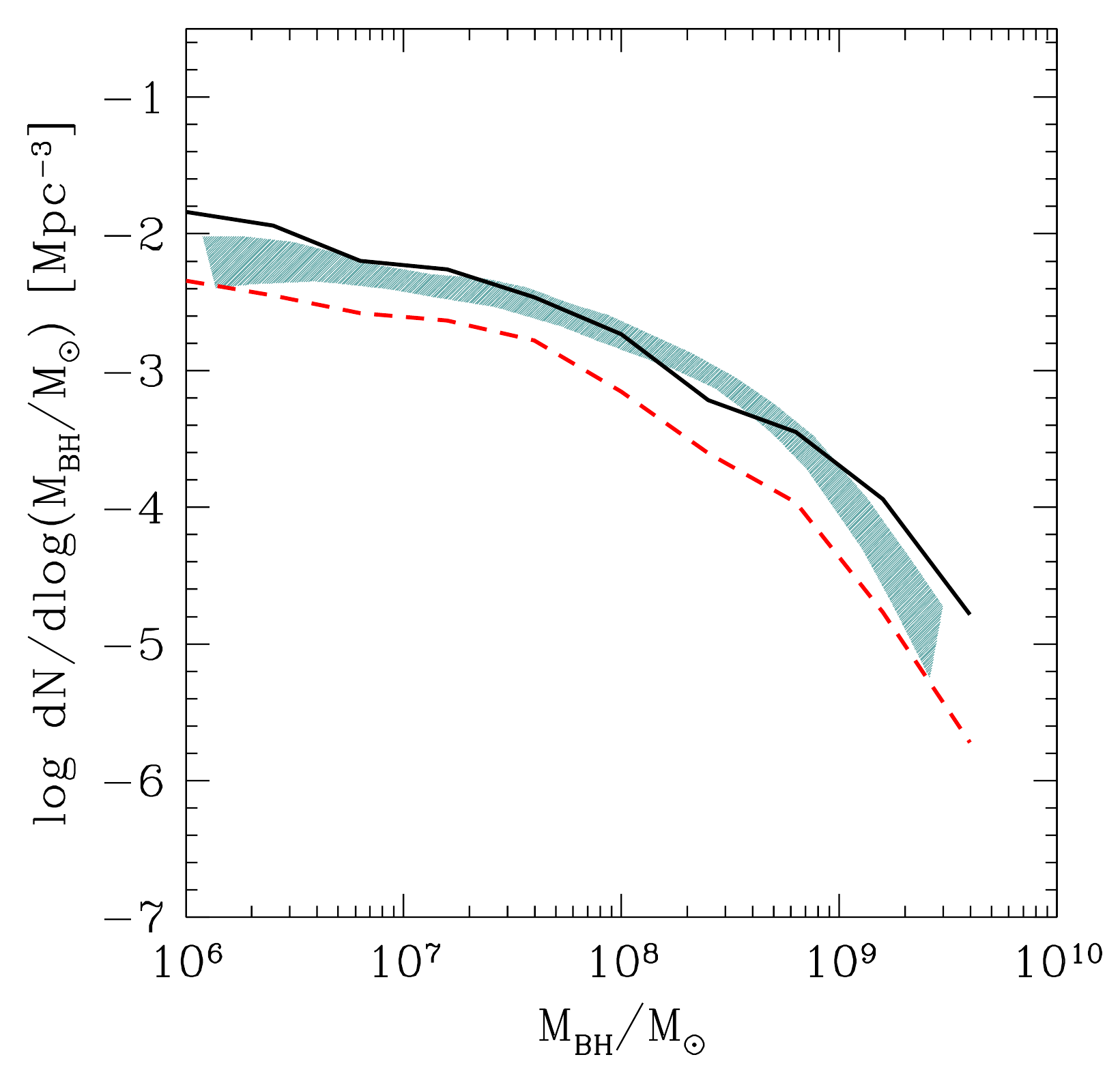}}}
\hspace{-0.2cm}
\scalebox{0.31}[0.31]{\rotatebox{-0}{\includegraphics{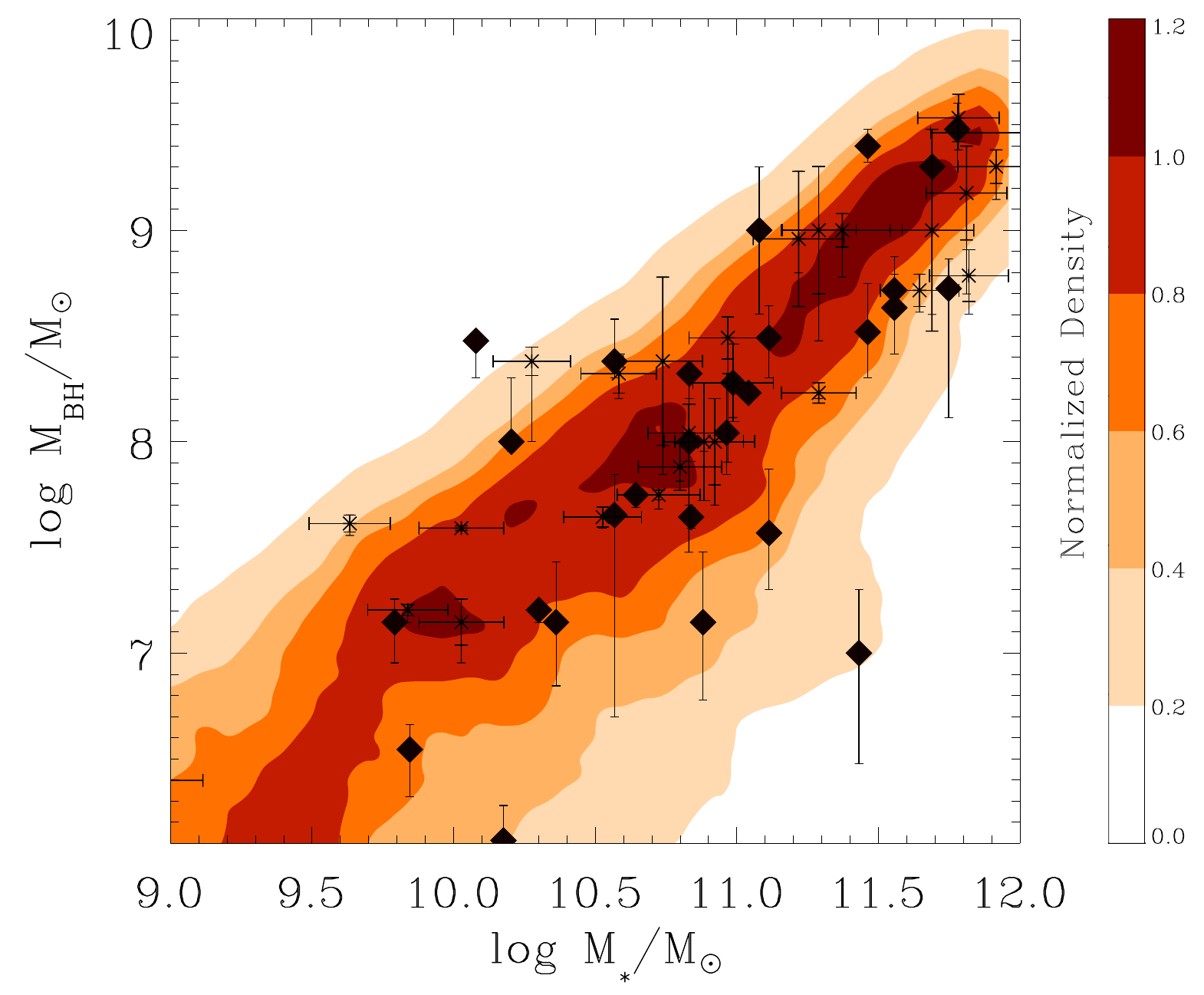}}}
\vspace{-0.cm }
 {\footnotesize 
Fig. 2a. The BH  mass function at $z<0.5$ built-up by  interaction-driven accretion (solid line) and by BH accretion due to disc instabilities (dashed line).  The shaded region defines the spread in observational estimates obtained using different methods, as compiled by Shankar, Weinberg, 
Miralda-Escud\'e (2009).\newline
Fig. 2b. The local relation between the BH mass and the total stellar content of galaxies is compared with data. 
Data points represent the observed local relation from  H\"aring and Rix (2004, diamonds), and 
Marconi \& Hunt (2003, squares, here $M_*$ is derived using the best-fitting virial relation of Cappellari et al. 2006); the colour code represents the fraction objects as a function of $M_{BH}$ for any given value of $M_*$, as indicated by the bar. 
}
\vspace{0.2cm }

In fig. 2a, the contributions from interaction-driven accretion and 
disc instabilities are also displayed, to  show how the former constitutes the dominant mode of BH growth, while fig. 2b illustrates the scatter  
that characterized the relation when the total stellar mass of the host galaxy is considered, due to the combined effect of the growth histories 
shown in fig. 1. However, selecting final galaxies with a specific dominant stellar component corresponds to selecting  growth histories 
with specific characteristics, yielding different relations between the BH mass and the selected stellar component.
In the next section, we shall dissect the stellar mass of host galaxies 
in the different components discussed above, to pin down the main source of the scatter in fig. 1b in relation to the growth histories, and to investigate the role played by the different physical processes described in this section to the building up of the BH mass. 
\section{Results}

Here we present the predicted correlation between the  BH mass and the different galaxy components. We shall compare our results with the observations presented in Kormendy \& Ho (2013), who collected available measurements of such quantities for a sample of 45 elliptical galaxies and 43 spiral and S0 galaxies; the latter include 21 galaxies with classical bulge and 22 with pseudo-bulges. The BH mass measurements are based on different methods:  stellar dynamics, CO molecular gas disc dynamics, maser disc dynamics, ionized gas dynamics. The latter, however, have been considered only when the observations took into account broad emission line widths (see above authors for details), and this contributes to 
  the larger normalization with respect to early works (see, e.g., Marconi \& Hunt 2003).  It is important to note that, although the sample represents a substantial collection of updated measurements, it does not constitute an homogeneous sample with well 
  defined selection critera, so that in the following we shall not attempt to perform statistical analysis concerning the relative abundance of BH found in the different galactic components (for the AGN luminosity functions,  BH mass function, and the B/T distributions predicted by our model see Menci et al. 2014).  We shall instead compare the data with the relation between $M_{BH}$ and the stellar mass of different components resulting in our model. 

The observed  $M_{BH}-M_{*,B}$ correlation between the BH mass and the mass of the classical bulge component is shown in fig. 3 and 
compared with the prediction of our model for the same component (built up by major mergers, sect. 2.2, point ii), for galaxies in the same magnitude range $-24\leq M_K\leq -21$  spanned by the observations. 
Within a standard deviation, the distribution we obtain is consistent with a single linear dependence in the log$\,M_{BH}$-log$\,M_{*,B}$ plane. 
The slope, the normalization, and the scatter of the model predictions are consistent with the observations. Indeed, a linear regression analysis of the predicted log$\,M_{BH}$-log$\,M_{*,B}$ distribution yields $log M_{BH}/M_{\odot}=(1.15\pm 0.1)\,log\,(M_{*.B}/M_{\odot})-3.84\pm 1$,  remarkably close to that derived by  Kormendy \& Ho 
(2013) for the data sample $log M_{BH}/M_{\odot}=(1.16\pm 0.08)\,log\,(M_{*.B}/M_{\odot})-4.1\pm 1$, but flatter than the intrinsic relation recently proposed by Shankar et al. (2016) on the basis of their analysis of the selection bias;  this is a non-linear relation, with a logarithmic slope $\approx  1.9$ and a  normalization close to that in Kormendy \& Ho (2013) at the large-mass end ($M_{*,B}\approx 10^{11.5}\,M_{\odot}$) of 
the $M_{BH}-M_{*,B}$ relation.  The scatter we obtain (0.4 dex, mildly dependent on the bulge mass, see the top histogram in fig. 3) is slightly larger than the intrinsic 
 scatter of the data sample $\approx 0.3$ dex (see Kormendy \& Ho 2013). 
Note that the exact value of the normalization in the predicted log $M_{BH}$-log $M_{*,B}$ relation, although similar to the  value measured by Kormendy \& Ho (2013), is not a true prediction of our model; in fact, it depends on the assumed ratio of BH accretion to nuclear star formation, a free quantity for which we adopted the fiducial value $1/4$ (sect. 2.1 and eq. 1). 

 The almost linear behaviour constitutes a natural outcome in a scenario where the interaction-driven accretion stimulates both nuclear star formation and BH accretion, as already shown in early SAMs with tunable accretion efficiency (Kauffmann \& Haenhelt 2000; Cattaneo et al. 2005; Croton et al. 2006).  However, approximately linear log $M_{BH}$-log $M_{*,B}$ relation also arise in several hydrodynamical simulations (see, e.g., Di Matteo et al. 2005; for recent cosmological simulations see Sijacki et al. 2015; Volonteri et al. 2016) which assume BHs to grow through Bondi accretion. In such simulations the fast runaway in the BH mass produced by the quadratic dependence on $M_{BH}$ of the accretion rate is counteracted by strong 
 feedback to produce  $M_{BH}$-$M_{*,B}$ relations close to the observed one, with a logarithmic slope that ranges from 0.75 to 1.25, also depending on the adopted method to measure the bulge mass in the simulations.  Thus, the slope in the $M_{BH}$-$M_{*,B}$ relation does not constitute a
sensible probe to pin down the physical origin of the observed $M_{BH}$-$M_{*,B}$ relation. In this context, studying how the  
the growth of BHs and of the different stellar components are connected may help to provide constraints on the physical origin of the BH-galaxy co-evolution. 

To this aim, we start studying the paths followed by the galaxy progenitors in the building up of the local relations. For the case of the 
$M_{BH}$-$M_{*,B}$ relation, these
 are presented in fig. 3 (bottom panel). This shows the time evolution of $M_{*,B}$ and $M_{BH}$ contributed by all progenitors of a few selected final galaxies, from $z=8.3$ to the final position in the log$\,M_{BH}$-log$\,M_{*,B}$ relation shown in the contour plot on the top. The paths have been chosen as 
 to end up in different regions of the final $\,M_{BH}$-$\,M_{*,B}$ relation corresponding to final galaxies 
 with $10^9\,M_{\odot}\leq M_{*,B}\leq 10^{11}\,M_{\odot}$ (note that the final part of the trajectories is covered by the top-most plane). In the figure, the circles mark the onset of a BH accretion phase (AGN), and their sizes corresponds to the Eddington ratio (see caption and legend). 
\begin{center}
\vspace{-0.3cm}
\hspace{-0.4cm}
\scalebox{0.48}[0.48]{\rotatebox{-0}{\includegraphics{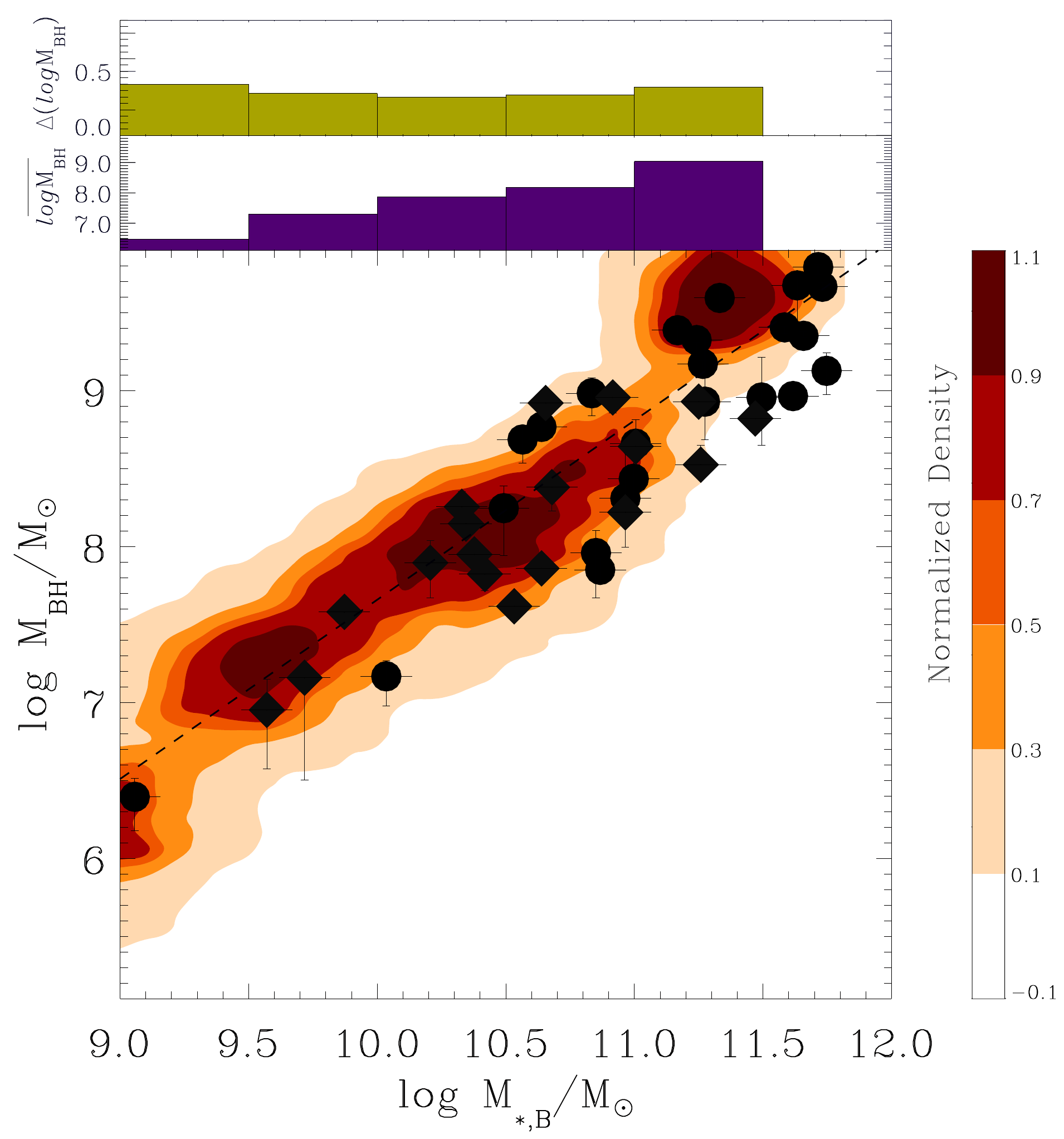}}}
\hspace{0.4cm}
\scalebox{0.42}[0.42]{\rotatebox{-0}{\includegraphics{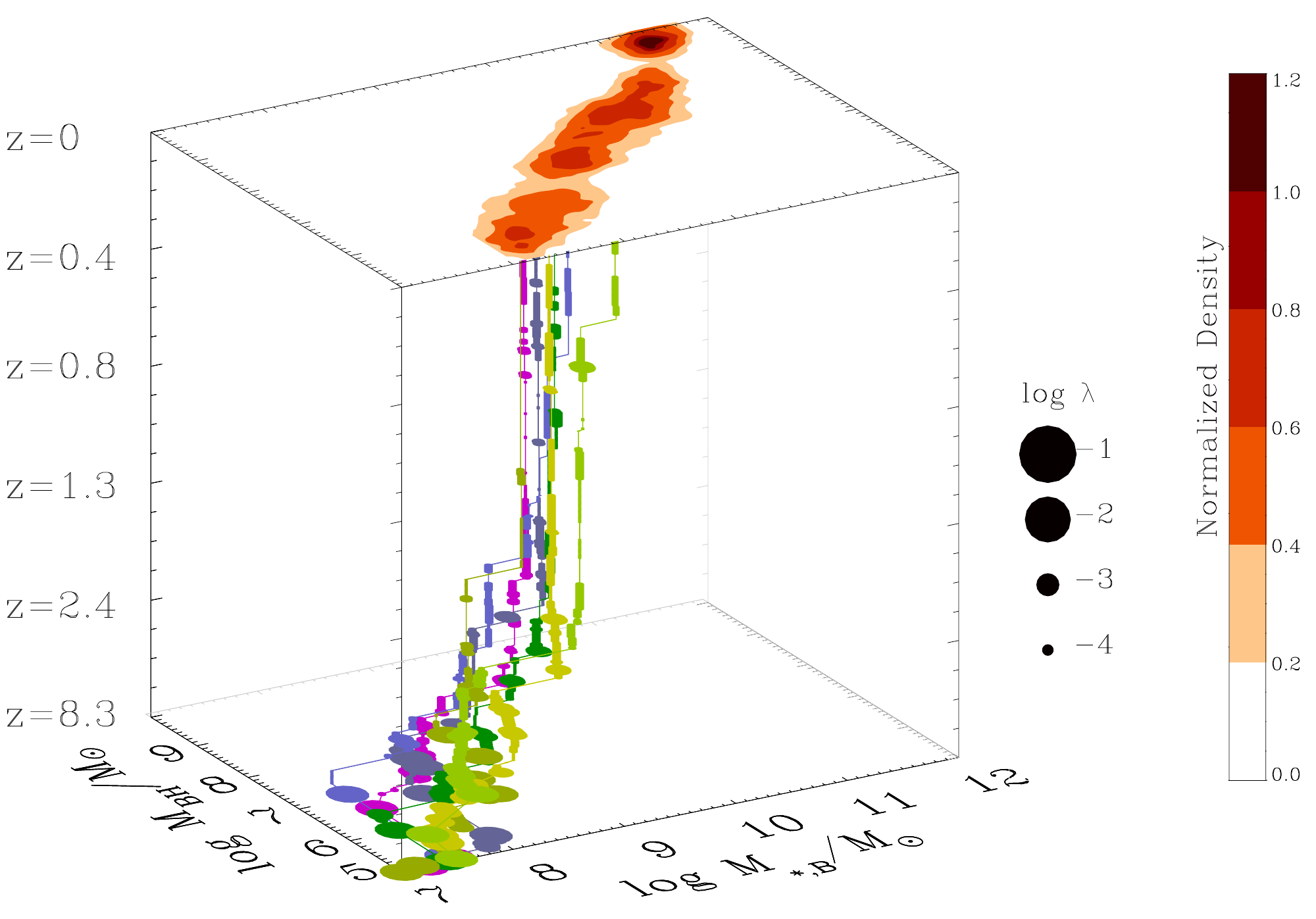}}}
\end{center}
\vspace{-0.1cm }
 {\footnotesize 
Fig. 3. Upper Panel.  The contour plot shows the predicted local relation $M_{BH}-M_{*,B}$ between the BH mass and the classical bulge mass. The contours indicate the number of galaxies in a $M_{BH}-M_{*,B}$ bin, normalized to the total  number at fixed $M_{*,B}$ as shown in the color bar. We compare with the data reported in Kormendy \& Ho (2013) for ellipticals  (circles) and classical bulges (diamonds) in spiral galaxies. 
 The dashed line in the contour plot shows the   fit $log M_{BH}/M_{\odot}=1.15\,log\,(M_{*.B}/M_{\odot})-3.84$, while the histograms on the top show the average value $\overline {log M_{BH}}$ and the scatter $\Delta {log M_{BH}}$ of the distribution of $log M_{BH}/M_{\odot}$  in different bins of bulge stellar mass. \\
Lower Panel. Each colored path shows the time evolution of the BH mass $M_{BH}(t)$ and bulge mass $M_{*B}(t)$ contributed by all progenitors of  a few  selected final galaxies. The circles show the onset of an active AGN phase; their size corresponds to the Eddington ratio $\lambda\equiv \dot M_{BH}/\dot M_{Edd}$ as shown in the legend.
\vspace{0.3cm}}

Interactions are the leading trigger for BH accretion in our model (see sec. 2.1-2.3 and fig. 2), so they drive the growth of both the BH and bulge mass. Since in CDM models interactions between comparable clumps (those most effective in triggering starbursts and BH growth, see eq. 1) take place  mainly at high redshifts (see, e.g., Zhao et al. 2003), the growth histories are characterized by a rapid increase in $M_{BH}$ and $M_{*,B}$ in a short interval of cosmic time ($z\gtrsim 2.5$), after which most of the gas has been turned into stars, leaving passively evolving bulges where only occasional events will trigger minor BH accretion. The short time lapse when the same physical process triggers the growth of both $M_{BH}$ and $M_{*,B}$ results in a final tight correlation as we show later. 

A different situation occurs when pseudo-bulges are considered. In fig. 4 we show the distribution of BH masses as a function of the 
pseudo-bulge stellar mass $M_{*,PB}$ of the host, for the model galaxies where such stellar component dominates over the classical bulge component,  to compare with the  data obtained under similar selection criteria, see Kormendy \& Ho 2013). 

\begin{center}
\vspace{-0.1cm}
\hspace{0.cm}
\scalebox{0.47}[0.47]{\rotatebox{-0}{\includegraphics{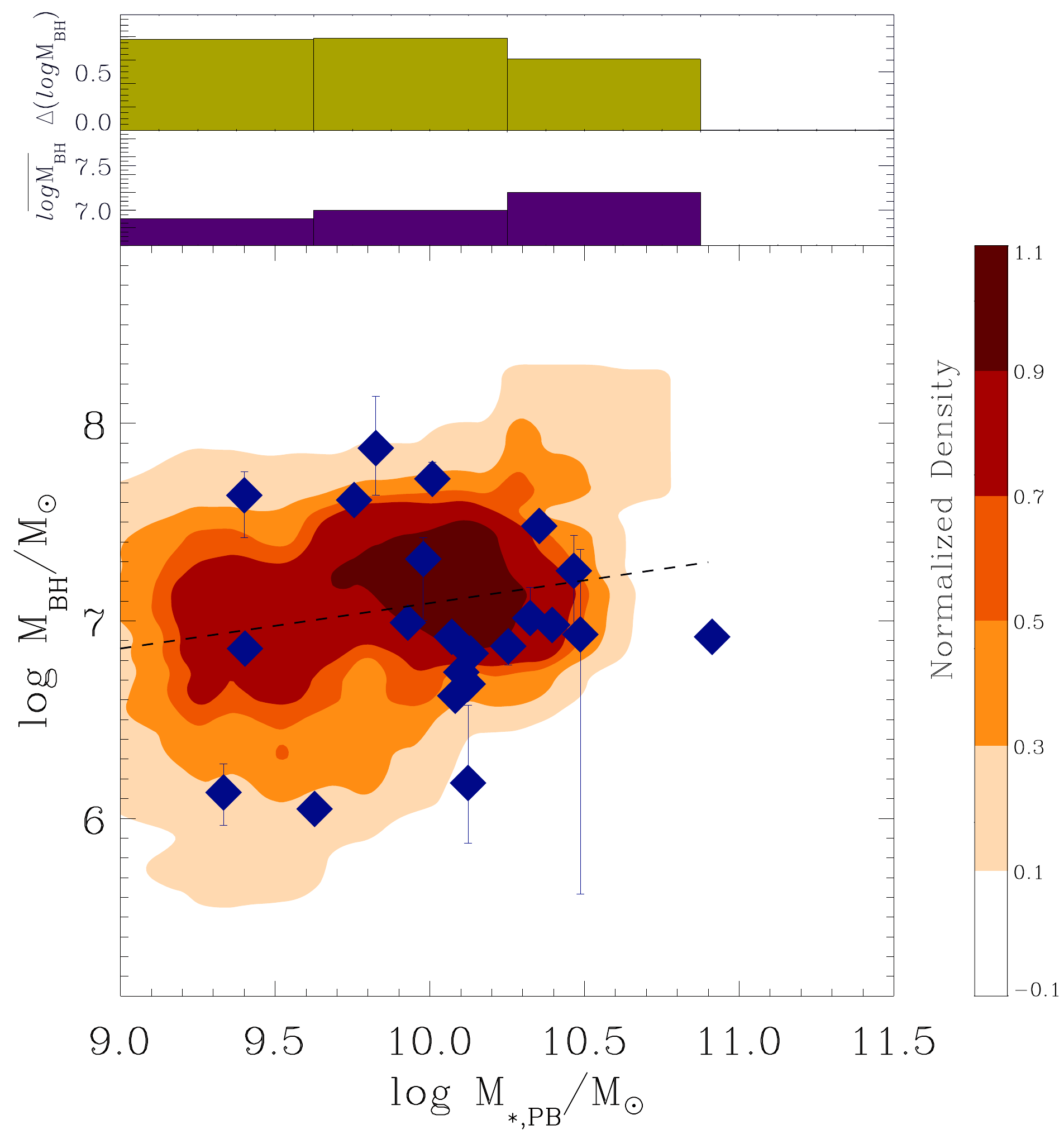}}}
\hspace{0.4cm}
\scalebox{0.44}[0.44]{\rotatebox{-0}{\includegraphics{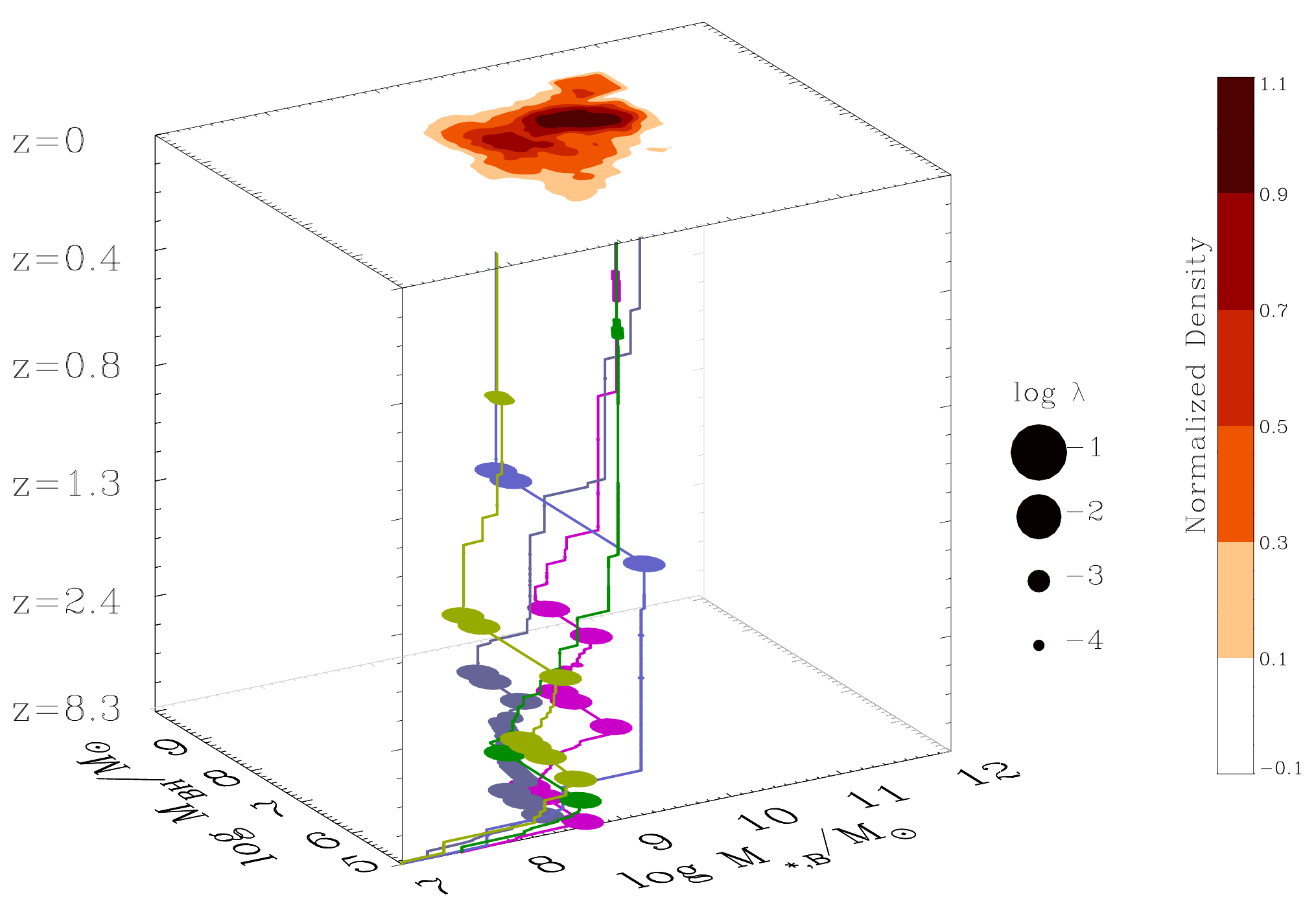}}}
\end{center}
\vspace{-0.4cm }
 {\footnotesize 
Fig. 4. As in fig. 1, but here we consider the relation between the BH mass $M_{BH}$ and the mass of the pseudo-bulge component $M_{*,PB}$, for  galaxies where the latter dominates over the classical bulge component. 
The dashed line in the top panel corresponds to a linear fit $log M_{BH}/M_{\odot}=0.23\,log\,(M_{*.B}/M_{\odot})+4.79$.
\vspace{0.3cm}}

 Compared to the previous case in fig. 3, a much weaker correlation (with a logarithmic slope 0.23) and a larger scatter $\sim 0.7$ are obtained.
In our model, pseudo-bulges  originate from the loss of angular momentum due to disc instabilities (sect. 2.2, point iii), so that the normalization of the computed magnitude depends on the parameter $A_*$ determining the normalization of such a star formation mode (sect. 2.2); the latter, in turn, depends on the density profile assumed for the outer disc region. Following our discussion in Sect. 2.2, for an exponential profile an horizontal shift up to 0.5 can affect the $M_K$ of the computed galaxies. However, this does not affect our main point here, namely, that the appreciably larger dispersion of the distribution compared to that obtained for classical bulges  (as shown by the larger scatter, shown in the top histogram of fig. 4), in agreement with the observed scatter shown by the data in  
Kormendy \& Ho (2013). This is because disc instabilities - building up the pseudobulge component  in our model - provide only a  minor contribution in the growth of SMBHs (see fig. 2a), so that the growth of pseudo-bulges is in practice uncorrelated 
to that of (most of the) BH mass, as shown by the  paths in the bottom panel of fig. 4. The same growth paths also show that, since disc instabilities require large gas fractions (eq. 3), the growth of a substantial pseudo-bulge must take place along those histories characterized by a low number of major merging at high redshifts, in order to avoid the  consumption of gas occurring in major mergers at early times. Thus, the paths leading to BH in galaxies with a dominant pseudo-bulge are characterized by a) BH growth largely uncorrelated to the growth of the pseudo-bulge, and b) the formation of the stellar mass $\Delta M_*$  spread out over a much wider lap of cosmic times compared to classical bulges, with a much smaller number of merging events $N$. {The combination of a) and b) result into a larger dispersion for log $M_{BH}$ (for a more extended interpretation, see the Discussion in the 
next section)}. 

Note that switching off interactions as BH growth mechanism would indeed 
restore an almost linear $M_{BH}-M_{*,PB}$ correlation (as shown in Menci et al. 2014; see also Angles-Alcazar et al. 2016), although this would result 
into an evolution of the AGN luminosity functions not matching the observations at both the faint and the bright end.

An  interesting consequence of the  interplay between the different star formation modes and the growth of BHs resulting from the modelling in Sect. 2 is the predicted  weak correlation between BH mass and the stellar disc component, as is shown in fig. 5. This shows the distribution of BH masses as a function of the $K$-band 
magnitude $M_{K,disc}$ (to directly compare with available data) of the disc stellar component of our simulated galaxies.  The relation we find is characterized by a large scatter that is broadly consistent with the dispersion of the data points. Again, the origin of the model results can be traced back to the growth of the BHs in relation to the star formation history of the disc component, shown in the bottom panel. While BHs growth is mainly triggered by interactions which rapidly increase the BH mass in a very short time scale $\sim \tau_d$ such a process is not directly correlated with the growth of the stellar disc component which is due to the quiescent gas conversion (point i) in sect. 2.2) taking place over a much longer time scale $\sim $ Gyr. The latter is responsible for the continuous growth of the stellar mass (at fixed BH mass) shown by the paths in the bottom panel of fig. 5. 

Note that such an effect for galaxies with a prominent disc component is deeply rooted in the cosmological properties of galaxy formation. 
In fact, such galaxies must have formed in lower density regions (compared to bulge-dominated systems), where the milder interaction rate
 allows the persistence of the disc. This results into lower star formation rates at high redshifts $z>4$, leaving a 
 larger reservoir of cold gas available for accretion at later cosmic times, thus allowing for a prolonged accretion phase 
at lower redshifts $z<2$.

\begin{center}
\vspace{-0.3cm}
\hspace{0cm}
\scalebox{0.47}[0.47]{\rotatebox{-0}{\includegraphics{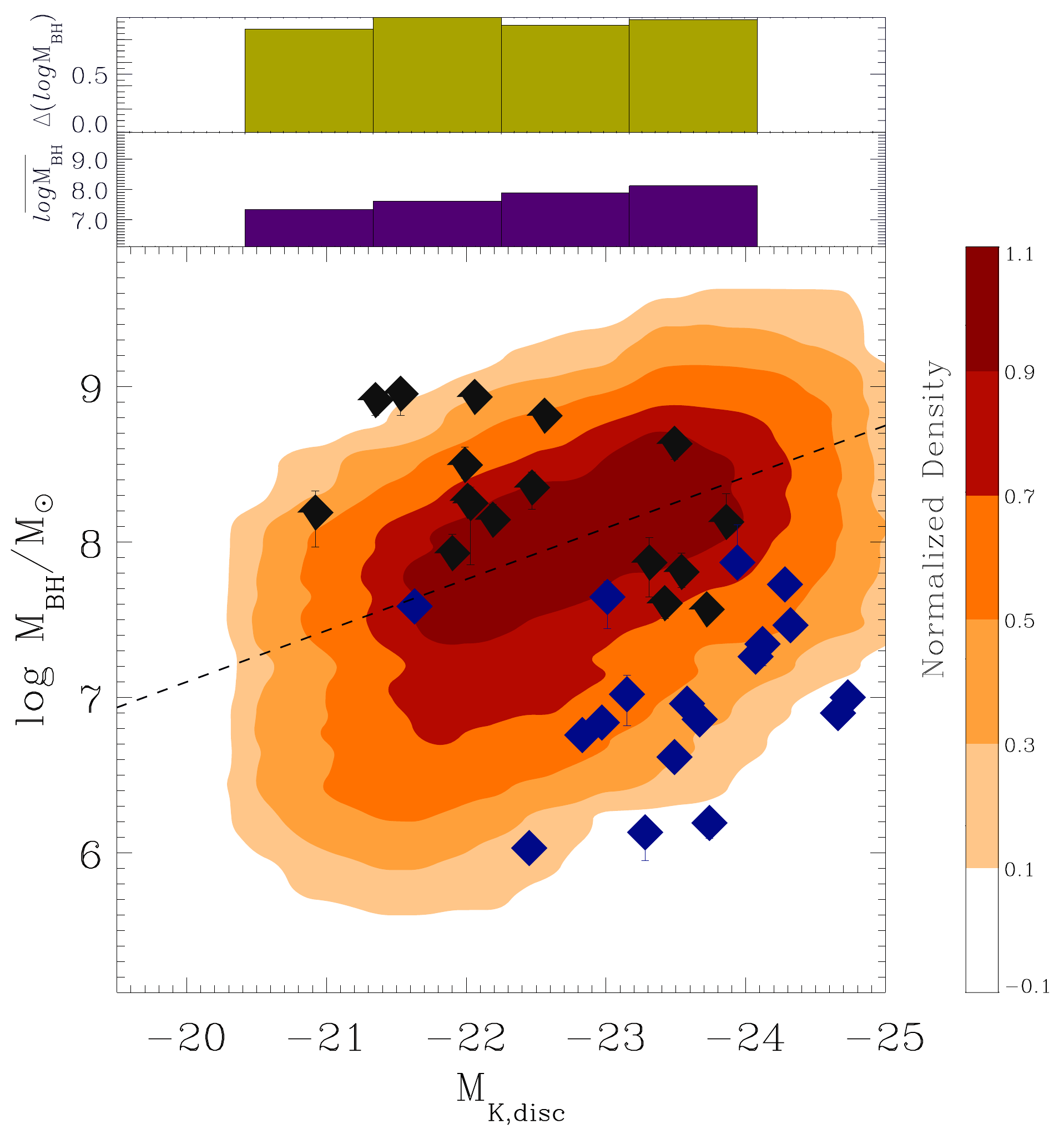}}}
\hspace{0.4cm}
\scalebox{0.44}[0.44]{\rotatebox{-0}{\includegraphics{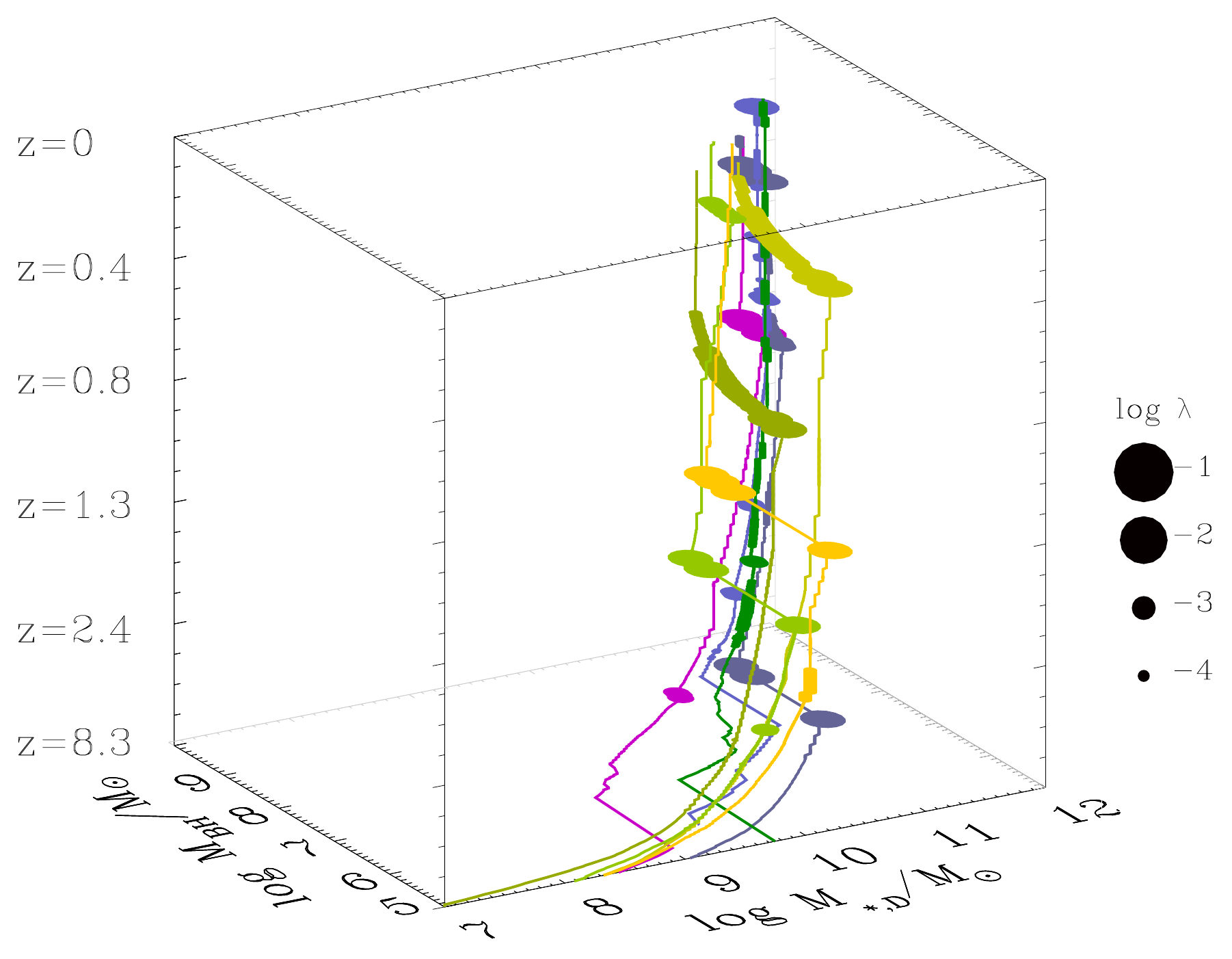}}}
\end{center}
\vspace{-0.4cm }
 {\footnotesize 
Fig. 5. Upper  Panel. The predicted distribution of galaxies in the $M_{BH}-M_{K,disc}$ plane from our SAM (contour plot), where $M_{K,disc}$ is the $K$-band magnitude of the stellar disc. The data are from Kormendy \& Ho (2013): black diamonds refer to spiral galaxies with classical bulges, while blue diamonds  correspond to galaxies with a pseudo-bulge.
 The dashed line in the contour plot shows the linear  fit $log M_{BH}/M_{\odot}=-0.264\,log\,(M_{K,disc}/M_{\odot})+1.97$, while the histograms on the top show the average value $\overline {log M_{BH}}$ and the scatter $\Delta {log M_{BH}}$ of the distribution of $log M_{BH}/M_{\odot}$. 

Lower Panel. Each colored path shows the time evolution of the BH mass $M_{BH}(t)$ and disc mass $M_{*,D}(t)$ contributed by all progenitors of a few  selected  galaxies. The circles show the onset of an active AGN phase; their size corresponds to the logarithm of the Eddington ratio $\lambda\equiv \dot M_{BH}/\dot M_{Edd}$ as shown in the legend. 
\vspace{0.2cm}}

An extreme instance of the above behaviour is constituted by bulgeless galaxies. These have been selected from the Mock catalogue generated by our SAM as those characterized by bulge-to-total ratio $B/T\leq 0.1$. The corresponding distribution of the hosted BH masses as a function of the disc $K$-band magnitude   is shown in fig. 6, and compared with the data presented in Kormendy \& Ho (2013).  We note that the requirement of low $B/T\leq 0.1$ translates into
selecting model galaxies originating from those (rare)  histories characterized by the absence of major merging at large redshifts. In this case, the BH growth is due to the later fly-by or minor merging events that do not lead to bulge formation. In this case the BH growth is  is not only uncorrelated with the growth of the disc component, but is spread over a large range of cosmic times $z\lesssim 1$ since high redshift mergers are excluded by definition. The results is a highly dispersed $M_{BH}-M_K$ relation. Its real consistence with the observational distribution must await more homogeneous observational samples with larger statistics and - most important - well definite selection criteria. 

\begin{center}
\vspace{-0.2cm}
\hspace{-0.3cm}
\scalebox{0.44}[0.44]{\rotatebox{-0}{\includegraphics{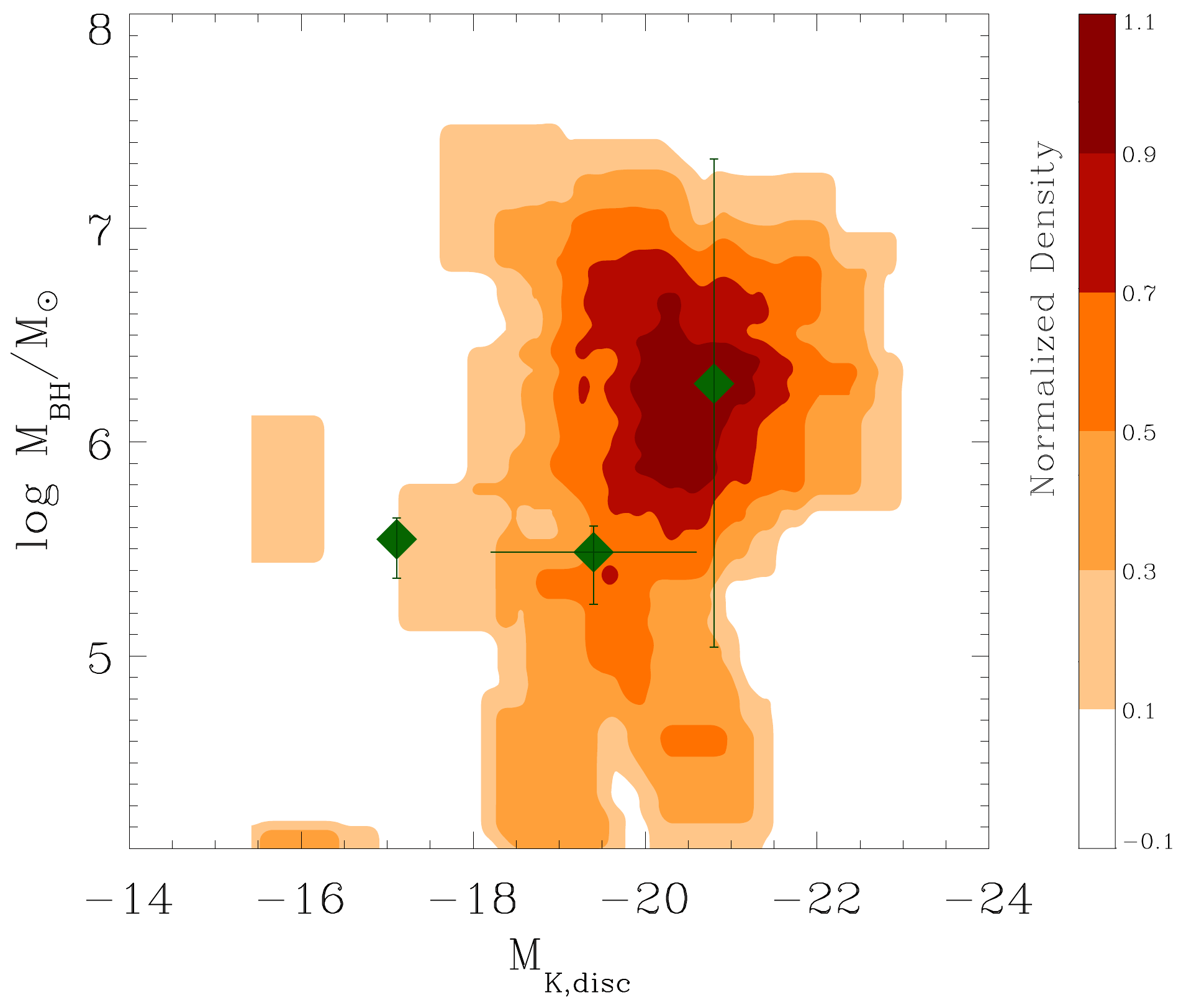}}}
\hspace{0.4cm}
\scalebox{0.42}[0.42]{\rotatebox{-0}{\includegraphics{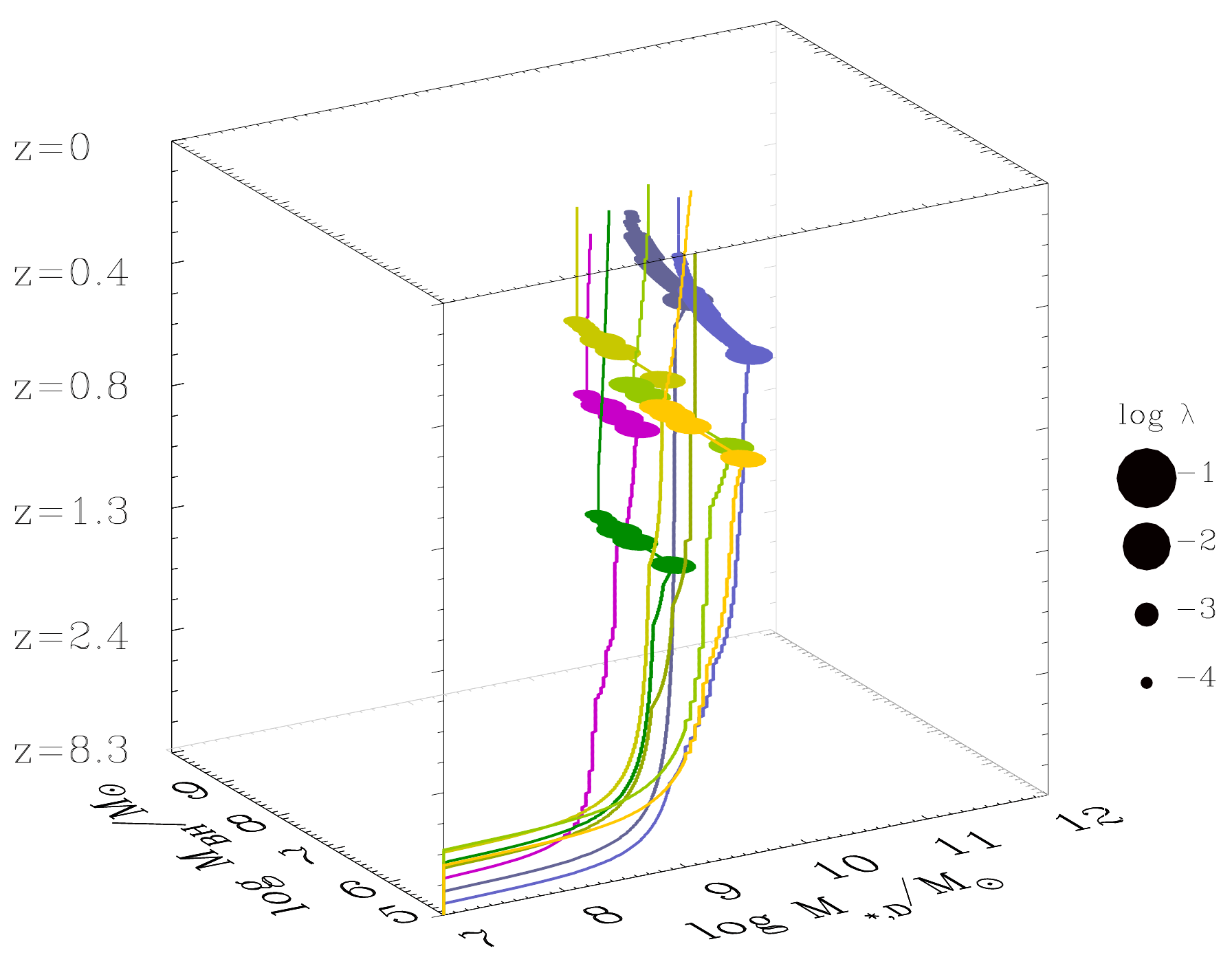}}}
\end{center}
\vspace{-0.4cm }
 {\footnotesize 
Fig. 6. As in fig. 5, but for  galaxies with no bulge. Data are taken from Kormendy \& Ho (2013), and are based on 
observations of  NGC4395 (Peterson et al. 2005), Pox53 (Thornton et al. 2008), and Henize 2-10 (Reines et al. 2011).
\vspace{0.2cm}}

The different behaviour of the BH mass distribution in bulges, pseudo-bulge and discs is compared and summarized in fig. 7. In the figure we plot the distribution function of $\Delta log M_{BH}$, the offset of the BH mass from the value where the $M_{BH}-M_*$ relation is peaked, for a reference stellar 
mass $M_{*}=2\cdot 10^{10}\,M_{\odot}$. Besides the appreciably larger dispersion obtained for the BH distribution in pseudo-bulges and discs compared to that in bulges, we note that for the former the distribution are skewed  toward lower BH masses. This might explain the  observational findings concerning the steepening of the global $M_{BH}-M_{sph}$ relation for small values of $M_{BH}$ reported by some authors (see Graham \& Scott 2015 and references therein). 

\begin{center}
\vspace{0.cm}
\hspace{-1.2cm}
\scalebox{0.3}[0.3]{\rotatebox{-0}{\includegraphics{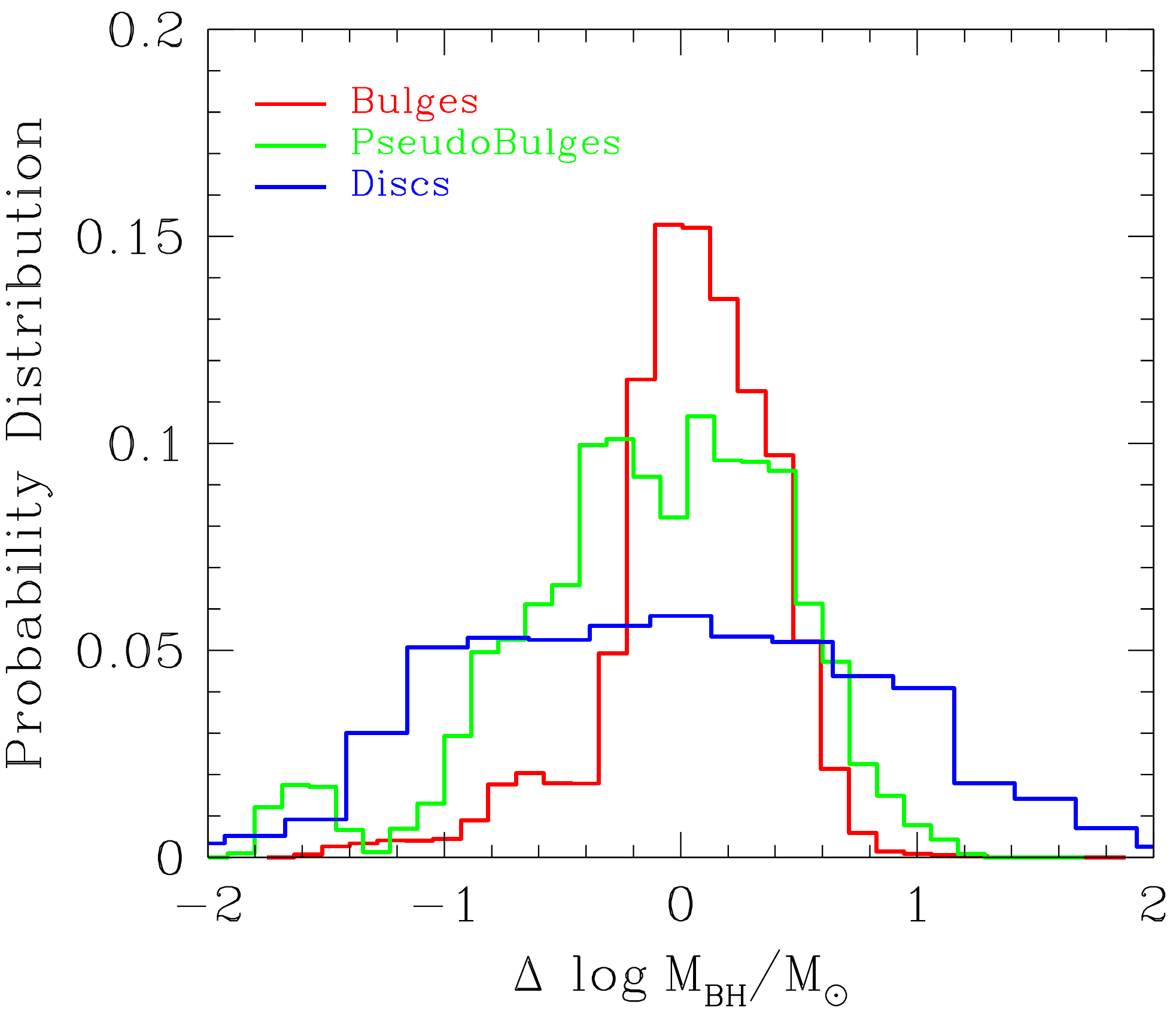}}}
\end{center}
\vspace{-0.4cm }
 {\footnotesize 
Fig. 7. The predicted distributions of $log M_{BH}$ around the peak value at a stellar mass $M_{*}=2\,10^{10}\,M_{\odot}$ is shown for 
bulges, pseudo-bulges and discs (red, green, and blue histogram, respectively). 
\vspace{-0.cm}}

\section{Discussion}

To better understand the origin of the small scatter in the relation in fig. 3  compared to that in figs. 4 and 5, we considered a simplified rendition of the processes considered in our SAM. Specifically, we consider a toy model (described in detail in Appendix), where the building up of a given stellar mass $\Delta M_*$ in a time $\Delta t$ is contributed by: i) the merging of galaxies containing stellar and BH masses; 
ii)  a term {\it correlated} to the BH growth,  corresponding to impulsive star formation (starbursts)  proportional (though a constant $C$) to the BH accretion rate. Such a term mimics the effects of merging and of the interactions, which in our model trigger BH accretion proportional to the star bursts, see sect. 2; iii)  a  term {\it uncorrelated} to the BH growth, with a star formation rate described in terms of a stochastic variable $\epsilon$. This represents the growth of the stellar mass due to secular processes not related to the BH growth (sect. 2.2), with the stochastic term $\epsilon$ reflecting the different content of cold gas that can be found in galaxies due to to their different merging histories and formation times (e.g., for the quiescent star formation $\epsilon\propto M_c/\tau_d$, see point {\it i} in sect. 2.2) . The above toy model allows to compute analytically (Appendix A) how the variance $\sigma (log M_{BH,f})$ of the distribution of final logarithmic BH masses $log M_{BH,f}$ is related to the initial value $\sigma (log M_{BH,i})$  associated to the distribution of progenitor galaxies, after $N$ merging events within the interval $\Delta t$. Such a toy model extends the simple case explored by Peng (2007) which considered only the  effect of merging on the variance of the $M_{BH}$ distribution. In the limit of large number of merging events $N$ one obtains (see Appendix A)
\begin{equation}
\sigma (log M_{BHf})={\sigma (log M_{BH,i})\over \sqrt{N}}+{C\,\sigma (\epsilon)\,\Delta t\over N\overline {M_{BH,i}}}
\end{equation}
where $\overline {M_{BH,i}}$ denotes the average of the initial BH mass distribution, and $\sigma (\epsilon)$ is the variance 
generated by all star formation processes not correlated to the BH accretion. The first term in eq. 5 is the same obtained by Peng (2007) and describes the statistical effect of merging, which yields a fractional scatter decreasing  during the growth of the stellar mass as $1/\sqrt{N}$ simply due to Law of Large Numbers. The second term in eq. 4 arises only in presence of BH accretion ($C\neq 0$) . It vanishes for star formation entirely correlated with the BH growth ($\sigma(\epsilon)\rightarrow 0$),  and increases with the duration $\Delta t/N$ corresponding to the building up of the stellar mass
 $\Delta M_*$. 

 In the case of the build up of classical bulges through merging events shown in fig. 3, $N$ takes large values during a small time lapse $\Delta t$, since at high redshifts hierarchical models predict fast, numerous interactions to take place into a small time lapse $\Delta t$, corresponding to a fast build up of the stellar mass $\Delta M_*$. Following eq. 4, this implies that the effect of the term corresponding to growth of BH {\it uncorrelated} with the stellar mass is suppressed, and that the large number of merging events $N$ leads to a rapid decrease of the scatter in $log M_{BH}$, as in fact is shown by the paths in the bottom panel of fig. 3b. 

 Eq. 4 can be used for a more quantitative interpretation of the origin of the scatter in figs. 3-5. While, of course, the quantities entering the toy model 
in eq. 4 can be estimated only approximately (an exact computation of the BH accretion, star formation and merging processes requires the full semi-analytic computation), we compute a first-order estimate of the scatter expected from eq. 4 to provide 
a guideline to understand how the quantities characterizing the growth histories in the bottom panels of figs. 3-5 (number of mergers, time scales for the growth of BHa ad stellar masses) determine the final scatter in the correlation between $M_{BH}$ and the bulge, pseudo-bulge, and disc stellar masses. 

To this aim, we start from the case of classical bulges. We first compute  $N$ corresponding to 
the growth paths in fig. 3 (bottom panel). For $z\geq 3$, our SAM yields values ranging from $N\approx 7$ to $N\approx 10$ for major mergers ($\mu \gtrsim 0.2$) for final galaxies with $B/T>0.5$, thus showing a weak dependence of $N$ on the descendant stellar mass 
(the final mass of the growth path) also found in other works (see, e.g.,  Rodriguez-Gomez et al. 2015 for recent results from the Illustris simulation). 
Starting from a fractional scatter in BH masses 
$\sigma (log M_{BH,i})\approx 1$ at very high redshift $z\gtrsim 8$ (see bottom panel of fig. 4), the effect of merging events (first term in eq. 4) is to reduce the fractional scatter  to $1/\sqrt{N}\approx 0.3$, almost independently on the stellar mass. 
Then we can estimate the contribution to the scatter due to the second term on the r.h.s. of eq. 4, corresponding to the 
effect of star formation not correlated with the bulge growth (i.e., secular star formation). In fact, the quantities 
$C\,\sigma (\epsilon)/ \overline {M_{BH,i}}$ in eq. 4 can be recast as $C\,(\sigma (\epsilon)/\dot M_{*,MS})\,( \dot M_{*,MS}/ \overline {M_{BH,i}})$ where $\dot M_{*,MS}$ is  the star formation rate in the main sequence of the $\dot M_{*}-M_*$ plane (see, e.g., Rodighiero et al. 2011 and references therein; for a comparison of our SAM with the observed main sequence see Lamastra et al. 2013). 
Since $\sigma(\epsilon)$ is the scatter due to secular star formation, 
the ratio $\sigma (\epsilon)/\dot M_{*,MS}$ can be approximated by the fractional scatter in the star formation rate of main-sequence galaxies 
for which typical values  $\sigma (\epsilon)/\dot M_{*,MS}\approx 0.3$ are measured (almost independent on the stellar mass, at least up to $z\approx 3$ where measurements are available). 
Thus, the contribution to the scatter from secular star formation in eq. 4 can be recast as 
$3\,(C/N)\,(\dot M_{*,MS}/M_{\odot}\,yr^{-1})\,(10^8\,M_{\odot}/\overline{M_{BH,i}})\,\Delta t/{\rm Gyr}$. 
The relevant quantities can be estimated from the growth paths in the bottom panel of fig. 3: for initial BH masses $\overline {M_{BH,i}}\approx 5\cdot 10^6\,M_{\odot}$ and initial stellar masses $M_*\lesssim 10^8\,M_{\odot}$, the high-redshift main sequence star formation rate is 
$\dot M_{*,MS}\approx 0.1\,M_{\odot}\,yr^{-1}$ ( as results from our SAM, and also from extrapolating to low masses the main-sequence relation at $z\geq 2.5$ in Rodighiero et al. 2011). Adopting $C\approx 1/5$ (see Appendix) and $N\approx 10$ yields 
a contribution to the scatter from non-bursty star formation $\lesssim 0.06\,\Delta t/{\rm Gyr}$. Recalling that $\Delta t$ is the growth time of the 
BH and stellar mass, the fast growth phase of BHs and bulge masses at $z\gtrsim 2.5$ corresponds to $\Delta t\lesssim 2-3\,{\rm Gyr}$ (in the paths in the bottom panel of fig. 3 all events with $\lambda\gtrsim  10^{-3}$ dominating the accretion are concentrated in this epoch). 
Thus, the final contribution to the scatter from non-bursty star formation 
can be estimated as $\approx 0.1-0.2$. Note that such a contribution is - to first order - weakly dependent on the stellar mass, since  higher values of 
$\dot M_{*,MS}$ corresponding to larger stellar masses are compensated by larger progenitor BH masses $\overline{M_{BH,i}}$, while the 
number of mergers $N$ is approximately independent on the mass scale, as discussed above.

In sum, for the building up of classical bulges, we expect contributions from mergers and secular star formation summing up to 
$\approx 0.4$ and weakly dependent on the stellar mass, as indeed shown in the top histogram in fig. 3. The evolution of the scatter 
is initially driven by merging events which reduce it from 1 dex to 0.3 between $z\approx 8$ and $z\approx 3$; in the same period 
the accretion term in eq. 4 adds up only 0.1 dex. After $z\approx 2.5$ the scatter remains close to 0.4 dex. Such an evolution is consistent with 
the evolutionary paths in the bottom panel of fig. 3. 

We can now estimate the growth of the scatter expected for the relation between $M_{BH}$ and the  pseudo-bulge mass $M_{*,PB}$ shown in fig. 4, 
adopting the same line of arguments exposed above. In this case, the lower number of major merging events (averaging over the paths in fig. 4 yields values $N\approx 5$), and the longer timescales involved in BH formation continuing down to $z=1$ (corresponding to $\Delta t\approx 6$ Gyr) combine to provide a 
contribution to the fractional scatter $\approx 0.6-0.7$, close to the value obtained from the full semi-analytic model and  shown by the histogram on top of fig. 4. 
The same argument can explain the even larger scatter obtained for the relation between $M_{BH}$ and the disc magnitude
shown in eq. 5. The large values resulting from the model (see the top histogram in fig. 5) result from the longer time lapse $\Delta t$ now extending from $z\approx 8$ to $z\approx 0$, as shown by the histories in the bottom panel of fig. 5. 

\section{Conclusions}

 Motivated by recent observations indicating that  SMBHs correlate differently with the different stellar components of the host galaxies (see Kormendy \& Ho 2013), we use a semi-analytic model of galaxy formation to investigate the correlation of the different galaxy stellar components with the mass of the central SMBH (for a parallel study of the role of galaxy morphology in determining the spin distribution of SMBHs, see Sesana et al. 2016). The stellar mass in the disc, bulge, and pseudo-bulge of galaxies is related to quiescent star formation,  galaxy interactions, and the loss of angular momentum following disc instabilities, respectively. 

Although the model we adopt includes both interactions and disc instabilities in isolated galaxies as possible triggers for BH accretion, even  assuming the highest possible contribution fro disc instabilities the bulk of the final BH mass $M_{BH}$ is thus contributed by interactions. These are also responsible for the  growth of the mass $M_{*,B}$ of classical bulges and - in CDM models - are expected to take place at large redshift. Such a parallel evolution of  $M_{BH}$  and of $M_{*,B}$ within a short time lapse ($z\gtrsim 3$) yields the tight correlation between BHs and the bulge stellar mass. 
Our assumed modelling for disc instabilities naturally yields an appreciably larger scatter for the relation between $M_{BH}$ and the masses of pseudo-bulges $M_{*,PB}$. 
In fact, in our model the latter are entirely contributed by disc instabilities (a widely adopted view, implemented in other SAMs; see e.g., Fontanot et al. 2015a), which are favored in  gas-rich galaxies. The latter feature implies that the growth of a substantial pseudo-bulge must take place along those histories characterized by a low number of major merging at high redshifts in order to avoid the consumption of gas occurring in major mergers at early times. The longer time scales involved in the growth of pseudo-bulges, along with the decoupled growth of $M_{*,PB}$ and $M_{BH}$ 
(still mainly contributed by galaxy interactions) lead to the large scatter in the predicted $M_{BH}-M_{*,PB}$ relation. 
The longer time scales and decoupled growth with respect to the BHs also characterize the growth of the disc stellar component $M_{*,D}$ (star formation timescales $\sim 1$ Gyr), causing a  large scatter and a weak (or no)  correlation between $M_{BH}$ and $M_{*,D}$. The extreme 
instance of such a decoupled and delayed star formation with respect to the BH growth is constituted by bulgeless galaxies.
 
Our model does not yield a steepening of the slope of the {\it average} $M_{BH}-M_{*,B}$ at low BH masses reported by some authors  
  (see Graham \& Scott 2015) for $M_{BH}\lesssim 2\,\cdot 10^8\,M_{\odot}$, who proposed that such a break in the power-law 
  scaling of $M_{BH}$ with the bulge mass may indicate AGN  feedback as the main driver for the observed relations. 
 While on the observational side the presence of such a break in the power-law behaviour is still debated (as opposed to a larger scatter, see Kormendy \& Ho 2013), on the theoretical side  our results agree with the findings of other SAM (Fontanot et al. 2015b) in predicting that disc instabilities, AGN feedback, or other processes  considered in the present paper do not provide a steepening in the slope of the $M_{BH}-M_{sph}$ relation when the 
  peak BH mass (for any given $M_{sph}$) is considered. In fact, the model in Fontanot et al. (2015b) yields a steepening  for $M_{BH}\lesssim 10^8\,M_{\odot}$ only when it includes stellar feedback in star-forming bulges (a process not included in our model).  
  On the other hand, we predict an increase of the scatter for the correlation between  $M_{BH}$ and the  {\it total} stellar 
 mass $M_*$ for low-mass galaxies (see, e.g., Menci et al 2014) due to the larger contribution of the pseudo-bulge and disc components to the total stellar content in such objects; the skewness toward small BH masses of such a scatter due to the contribution of pseudo-bulges may explain the 
 steepening in the $M_{BH}-M_{sph}$ relation reported by Graham \& Scott (2015). 
  
Although in some phenomenological models a $M_{BH}-M_{*,B}$ relation  characterized by a single power-law can result from dry mergers driving BHs and their hosted galaxies towards a mean relation (Peng 2007; Hirschmann et al. 2010; Jahnke \& Maccio` 2011), in our model the BH growth is  
mainly contributed by BH accretion,  and the $M_{BH}-M_*$ relations results from the correlated BH and bulge growth from a common gas supply. The final $M_{BH}-M_{*,B}$ relation stems from the  physical relation between such an accretion and the building up of the bulge component in a short time lapse, followed by a phase ($z\lesssim 3$) in which the gas content of galaxies, their merging rate, and the BH accretion drop. A consistency check for such a framework is indeed constituted by  BH relation with the stellar mass which forms in the subsequent phase ($z\lesssim 3$), i.e., the disc and pseudo-bulge components: their weaker correlation with the BH mass indicate that the secular processes involved in their formation differ from the impulsive processes driving the build up of  SMBHs and the bulge in the earlier stages of galaxy evolution.

In our model, the establishment of the correlation between the BH mass and the different stellar components is determined by how the 
 growth of these components is connected to the common gas supply. Changing the  efficiency of AGN feedback has a minor effect on the predicted correlations presented in Sect. 3, as also obtained in other SAMs (see, Fontanot et al. 2015b and discussion therein). 
Of course, in our model, feedback affects both star formation and BH accretion, but it has a similar effect on both so that it does not constitute 
the driver of the $M_{BH}-M_*$ relations. Although the feedback origin of the connection between black holes and galaxies has been 
proposed or postulated in several analytical models (see, e.g., Silk \& Rees 1998; Fabian et al. 1999; King 2003, 2005, Whythe \& Loeb 2003), we stress that to establish the origin of the BH-galaxy correlation a detailed description of the relative growth of the BH and stellar components is needed. This,  
in turn, requires a detailed treatment of the several non-linear processes (gas cooling, disc formation, galaxy interactions) to connect them with the 
cosmological growth of the host galaxies, a task which is best accomplished through SAMs or N-body simulations. While in some simulations (see Di Matteo et al. 2005, 2008; Debuhr 2011; Choi et al. 2012; see also Costa et al. 2014) AGN feedback is required to establish the $M_{BH}-M_*$ relation, we stress that all such simulations assume Bondi accretion for the BH growth. Indeed, as already noticed by Angles-Alcazar et al. (2016), in any model where the dependence of the BH accretion rate on $M_{BH}$ is more than linear  (as in the Bondi accretion, scaling like $M_{BH}^2$) a divergent evolution of the 
BH mass would be obtained in the absence of a strong feedback. Thus, in such simulations, the need for feedback stems from the assumed accretion rate. In fact, in simulations where a milder dependence on $M_{BH}$ is assumed for BH accretion (see Angles-Alcazar et al. 2016), the establishment of the $M_{BH}-M_*$ relations is not originated by the feedback but rather by the relative growth of the stellar and BH components.
 
Our model provides in principle testable expectations for  the evolution of the relation of BH masses with the different stellar components. 
 Inspection of the paths $M_{BH}(t)-M_{*,B}(t)$ in the bottom panel of fig. 3 shows that for 
galaxies dominated by bulges the typical paths lie {\it above} the local relation, so we expect for such galaxies (and in general for massive galaxies 
formed from the collapse of biased regions of the density field) a steeper $M_{BH}-M_{*,B}$ relation at high redshifts (see Lamastra et al. 2010), 
consistently with available observations of QSOs extending up to $z\approx 4$ (Walter et al. 2004; Riechers et al. 2008). On the other hand, the evolutionary paths of $M_{BH}(t)-M_{*,B}(t)$ for the pseudo-bulges and disc components (bottom panels of figs. 4 and 5) show a mild trend toward lying {\it below} the local relation. Although at present observational limitations do not allow to explore the evolution of the relation of BH masses with the different stellar components, future measurements of such evolutionary paths for different stellar components will provide a stringent probe for the 
 role of the different BH feeding processes in determining the growth of BHs. 
 
Finally, we discuss the effects of relaxing the assumptions of our fiducial modelling on the relation between the BH mass and the different stellar components. The results presented above assume that classical bulges are formed from major mergers (see Sect. 2.2). However, some authors argue that the local frequency of spirals with low values of the $B/T$ ratio ($B/T\lesssim 0.2$) may indicate that secular processes can appreciably affect 
the formation of classical  bulges (see, e.g.,  Weinzirl,  Jogee,  Khochfar 2009;  Saha 2015), a possibility explored in some theoretical works based on SAMs (see Perry, Eke, Frenk 2009; De Lucia et al. 2011; see also Brenman et al. 2015). The effect of assuming such an hypothesis can be 
 explored in our model by assuming that all bulges are formed through the disc instability process (case iii in Sect. 2.2). The result of such an assumption would be a relation between BH mass and classical bulge  mass $M_{BH}-M_{*,B}$ similar to that shown in fig. 4, characterized by a large dispersion and very weak correlation. Although present data seem to exclude such a case (see, e.g., Kormendy \& Ho 2013), if future observational 
  improvements (or a refined treatment of observational biases, see Shankar et al. 2016) should lead to a revision of the present observational picture this would constitute a strong support for the disc instability origin of the majority of galaxy bulges. Another possible variant of our model is constituted by assuming disc instabilities to constitute the main contribution to the BH accretion.  
  Within our physical modelling of disc instabilities (sect. 2), such a feeding mode can indeed be comparable to galaxy interactions for intermediate AGN luminosity  (bolometric luminosity $10^{44}\lesssim L_{Bol}/{\rm erg\,s^{-1}}\lesssim 10^{46}$) at $z\approx 1.5-2.5$  (see Menci et al. 2014; Gatti et al. 2015). In the extreme case of a dominance of disc instability as BH accretion triggers in the above redshift range,  we expect a decrease of the dispersion in the correlation $M_{BH}-M_{*,PB}$ between the BH mass and the pseudo-bulge mass  at $z\approx 1-2$ since the growth of both quantities would be driven by the same physical process: at the same time, 
  we would expect a larger dispersion in the relation of BH masses with the masses of classical bulges. In the future, 
measurements of such a correlation in the above redshift interval will contribute to probe the relative role of interactions and disc instabilities in triggering the BH accretion, in the regimes where  disc instability may provide their maximal contribution. 

Finally, we note that additional mechanisms not implemented in our SAM may play a role in BH feeding at high redshifts, like violent disc instabilities (Bournaud et al. 2011) 
or even  direct formation of a compact bulge from the cosmic web (Dubois et al. 2012). We plan to investigate the effects of such mechanisms 
in future works. 

\begin{acknowledgements}
This work was partially supported by PRIN INAF 2011. We thank the referee for constructive comments that helped to improve the manuscript.\\ 
\end{acknowledgements}

\newpage
\section*{APPENDIX A}
\setcounter{equation}{0} \renewcommand{\theequation}{A-\arabic{equation}}
We consider the toy model introduced in Sect. 4, which is meant to provide a simplified analytical rendition of the growth of the stellar and 
BH masses of our SAM. In the toy model, galaxies are characterized by identical initial stellar masses $M_{*,i}$ each hosting a BH mass $M_{BH,i}$ uncorrelated with their host mass. A given, fixed stellar mass growth $\Delta M_*$ in a time interval $\Delta t$ is contributed by:
\begin{equation}
\Delta M_*=\Sigma_i^N M_{*,i}+\Delta M_{*,bursts}+\epsilon\,\Delta t.
\end{equation}
The first term corresponds to the  merging of $N$ galaxies  with initial stellar mass $M_{*,i}$; the second to the growth due to starbursts, 
the component {\it connected} to the BH accretion in our SAM; the third term corresponds to the  conversion of gas into stars {\it not correlated} to the BH feeding,  with the stochastic variable $\epsilon$ providing a simplified description of the distribution of star formation rate contributed by the secular processes (e.g., the quiescent star formation rate $M_c/\tau_d$ discussed in sect. 2.1). 

While the third term is taken to be {\it uncorrelated} to the BH growth, we assume the starburst term to be {\it correlated} to the BH accreted mass $\Delta M_{BH}=C\,\Delta M_{*,bursts}$; this mimics the description of interaction-driven BH feeding and starburst of our full SAM, described in Sect. 2. Note that, physically, the constant $C$ relating the BH accretion rate to the starburst takes values smaller than one (typical values in local ULRIGS values $C\approx 1/5-1/10$ are observed, Sanders \& Mirabel 1996). The limit $C\rightarrow 0$ corresponds to vanishing BH accretion. 
 
The accreted BH mass increment is the difference between the final BH mass and the sum of the initial mass of the BH hosted in the progenitor galaxies. 
Thus
$\Delta M_{BH}=C\,\Delta M_{*,burst}=M_{BH,f}-\Sigma_i^N M_{BH,i}$. Substituting the latter expression into  eq. A1  yields
\begin{equation}
M_{BH,f}=\Sigma_i^N M_{BH,i}+C\,\Delta M_*-C\,\Sigma_i^N M_{*,i}-C\,\epsilon\,\Delta t ~.
\end{equation}
The average final BH mass corresponding to eq. A-2 is then 
\begin{equation}
\overline{M_{BH,f}}=N\overline{ M_{BH,i}}+C\,\Delta M_*-C\,N\overline{M_{*,i}}-C\,\overline{\epsilon}\,\Delta t ~, 
\end{equation}
while the dispersion (square root of the variance) is 
\begin{equation}
\sigma(M_{BH,f})=\sqrt{N}\sigma( M_{BH,i})+C\,\sigma(\epsilon)\,\Delta t ~, 
\end{equation}
where $\sigma (\epsilon)$ is the variance 
generated by all star formation processes not correlated to the BH accretion. From eqs. A-3 and A-4 we obtain the fractional dispersion as
\begin{equation}
\sigma( log M_{BH,f})={\sigma(M_{BH,f})\over \overline{M_{BH,f}}}= {\sigma(M_{BH,i})+(1/\sqrt{N})\,C\,\sigma(\epsilon)\,\Delta t 
\over 
\sqrt{N}\,\overline{ M_{BH,i}}+(1/\sqrt{N})\overline{\Delta M_{BH}}.   }
\end{equation}
Taylor expansion of eq. A-5 for values $(1/\sqrt{N})\rightarrow 0$ yields
\begin{equation}
\sigma (log M_{BH},f)={\sigma (log M_{BH,i})\over \sqrt{N}}+{C\,\sigma (\epsilon)\,\Delta t\over N\overline {M_{BH,i}}}~.
\end{equation}
The first term which corresponds to the statistical suppression of the fractional dispersion 
pointed out by Peng (2007), which is recovered in the limit of vanishing BH accretion $C\rightarrow 0$. 
The second term in eq. A-6 is generated by the star formation 
uncorrelated to the BH growth.

\end{document}